\documentclass[acmsmall,table,xcdraw]{acmart}
\setcopyright{none} 
\def\BibTeX{{\rm B\kern-.05em{\sc i\kern-.025em b}\kern-.08emT\kern-.1667em\lower.7ex\hbox{E}\kern-.125emX}}

\usepackage[many]{tcolorbox}
\usepackage{url}
\usepackage{hyperref}
\usepackage{nicefrac}
\usepackage{siunitx}
\usepackage{array,framed}
\usepackage{pdfpages}
\usepackage{booktabs}
\usepackage{
  color,
  float,
  epsfig,
  graphics,
  graphicx,
  subcaption
}

\usepackage{amsfonts}
\usepackage{latexsym,fancyhdr}
\usepackage{enumerate}
\usepackage{algorithm,algorithmic}
\usepackage{xparse} 
\usepackage{xspace}
\usepackage{multirow}
\usepackage{csvsimple}
\usepackage{bbding}
\usepackage{pifont}
\usepackage{colortbl}
\usepackage{threeparttable}
\usepackage{longtable}
\usepackage{soul}
\newtheorem{definition}{Def}

\usepackage{
  tikz,
  pgfplots,
  pgfplotstable
}

\usetikzlibrary{
  shapes.geometric,
  arrows,
  external,
  pgfplots.groupplots,
  matrix
}

\usepackage{mathtools,}

\newcommand{\llamaseven}{\textsc{Llama-2-7b-chat}\xspace}
\newcommand{\llamatwo}{\textsc{Llama-2}\xspace}
\newcommand{\mistralseven}{\textsc{Mistral-7b-Instruct}\xspace}
\newcommand{\gemmanine}{\textsc{gemma-2-9b-it}\xspace}

\newcommand{\llamathreeit}{\textsc{Llama-3-8b-Instruct}\xspace}
\newcommand{\tool}{\textsc{D-LLM}\xspace}

\newcommand{\llamaguard}{\textsc{Llama-Guard-3}\xspace}
\newcommand{\jailbench}{\texttt{JailbreakBench}\xspace}

\begin{document}
%
% paper title
% Titles are generally capitalized except for words such as a, an, and, as,
% at, but, by, for, in, nor, of, on, or, the, to and up, which are usually
% not capitalized unless they are the first or last word of the title.
% Linebreaks \\ can be used within to get better formatting as desired.
% Do not put math or special symbols in the title.
\title{Model-Editing-Based Jailbreak against Safety-aligned Large
Language Models}

\author{Yuxi Li}
\affiliation{%
  \institution{Huazhong University of Science and Technology}
  \city{Wuhan}
  \country{China}
}
\email{yuxili@hust.edu.cn}

\author{Zhibo Zhang}
\affiliation{%
  \institution{Huazhong University of Science and Technology}
  \city{Wuhan}
  \country{China}
}
\email{zhangzhibom@hust.edu.cn}

\author{Kailong Wang}
\authornote{Corresponding Author.}
\affiliation{%
  \institution{Huazhong University of Science and Technology}
  \city{Wuhan}
  \country{China}
}
\email{wangkl@hust.edu.cn}

\author{Ling Shi}
\affiliation{%
  \institution{Nanyang Technological University}
  \city{Singapore}
  \country{Singapore}
}
\email{ling.shi@ntu.edu.sg}

\author{Haoyu Wang}
\orcid{0000-0003-1100-8633}
\affiliation{%
  \institution{Huazhong University of Science and Technology}
  \city{Wuhan}
  \country{China}
}
\email{haoyuwang@hust.edu.cn}

\begin{abstract}
Large Language Models (LLMs) have transformed numerous fields by enabling advanced natural language interactions but remain susceptible to critical vulnerabilities, particularly jailbreak attacks. Current jailbreak techniques, while effective, often depend on input modifications, making them detectable and limiting their stealth and scalability. This paper presents Targeted Model Editing (TME), a novel white-box approach that bypasses safety filters by minimally altering internal model structures while preserving the model's intended functionalities. TME identifies and removes safety-critical transformations (SCTs) embedded in model matrices, enabling malicious queries to bypass restrictions without input modifications. By analyzing distinct activation patterns between safe and unsafe queries, TME isolates and approximates SCTs through an optimization process. Implemented in the \tool framework, our method achieves an average Attack Success Rate (ASR) of 84.86\% on four mainstream open-source LLMs, maintaining high performance. Unlike existing methods, \tool eliminates the need for specific triggers or harmful response collections, offering a stealthier and more effective jailbreak strategy. This work reveals a covert and robust threat vector in LLM security and emphasizes the need for stronger safeguards in model safety alignment.

\textbf{\textcolor{red}{Warning:}} This paper includes examples of potentially harmful information solely for illustrative purposes. Readers are cautioned against misuse.%, and the content is intended only to highlight vulnerabilities and enhance security.
\end{abstract}

% make the title area
\maketitle

\section{Introduction}
Large Language Models (LLMs) have rapidly advanced in recent years, transforming various domains by enabling human-like interactions, generating coherent text, and performing complex tasks across industries. These models are powered by massive datasets and sophisticated neural architectures, enabling them to comprehend and generate natural language with remarkable precision. However, as LLMs continue to gain prominence, they also face critical threats to their reliability and robustness. These vulnerabilities, if exploited, can lead to misuse, including generating harmful content~\cite{zou2023GCGadvbench, shen2024donowcharacterizingevaluating, zhang2024SP, zhao2024weaktostrongjailbreakinglargelanguage, wang2024attngcgenhancingjailbreakingattacks}, leaking sensitive information~\cite{lukas2023analyzingleakagepersonallyidentifiable, zhang2024adanonymizerinteractivelynavigatingbalancing, qian2024deandeactivatingcoupledneurons, burgess2024privacyhardenedhallucinationresistantsyntheticdata}, or providing biased or misleading outputs~\cite{li2024drowzeemetamorphictestingfactconflicting, tang2024gendercarecomprehensiveframeworkassessing, pal-etal-2023-med, zhang2023sirenssongaiocean, huang2023surveyhallucinationlargelanguage, li2024glitchhunter, zhang2024glitchprober}. 
% \shil{Can add Nature Hallucination paper and change Drowzee to OOPSLA bitbex}
%Ensuring the security and robustness of LLMs has become an urgent concern, as they are increasingly integrated into systems that directly impact human decisions and interactions. Existing defenses, while evolving, have struggled to fully address the dynamic nature of the threats these models face.

One notable category of threat that has garnered significant attention is jailbreak attacks. These attacks exploit vulnerabilities in the safety and security mechanisms embedded within LLMs, circumventing the restrictions designed to enforce responsible behavior. These attacks can be broadly classified into two types: black-box and white-box approaches. Black-box attacks~\cite{Deng2024MASTERKEY, yu2024gptfuzzerredteaminglarge, deng2024pandorajailbreakgptsretrieval, liu2024autodangeneratingstealthyjailbreak} operate without knowledge of the model’s internal structure, relying on iterative adjustments  of input prompts based on the model’s outputs to elicit unintended or harmful behavior while avoiding detection. White-box attacks~\cite{zou2023GCGadvbench, guo2024coldattackjailbreakingllmsstealthiness, wallace2021universaladversarialtriggersattacking}, on the other hand, leverage access to the model’s architecture, parameters, or training data to craft more targeted jailbreak attempts. For example, backdoor injection for jailbreak attacks~\cite{rando2024universaljailbreakbackdoorspoisoned} could induce the model to produce target malicious responses. 
% Existing jailbreak techniques in the literature can be broadly categorized into two approaches: black-box and white-box attacks. Black-box attacks treat the LLM as an opaque entity, relying solely on the model's outputs to iteratively refine the input prompts. Attackers analyze the generated responses and systematically adjust their prompts to provoke unintended or malicious behavior while evading the model's safety filters. On the other hand, white-box attacks assume some level of access to the model's internal workings, such as its architecture, parameters, or training data. This additional knowledge allows attackers to craft more targeted and effective jailbreak attempts. For example, backdoor attacks could leverage fine-tuning techniques to inject carefully crafted malicious into the model, aiming to elicit targeted malicious responses. Regardless of the approach, the majority of existing jailbreak techniques fundamentally rely on optimization methods. Attackers iteratively refine their modifications based on the model's responses, employing various optimization algorithms to search for solutions that successfully bypass the safety measures.
%Regardless of whether these attacks follow a white-box or black-box approach, existing works in the literature fundamentally rely on optimization techniques \textbf{(on input prompts)}. More specifically, attackers refine their inputs iteratively by analyzing the outputs generated by the LLM, systematically adjusting their prompts to provoke unintended or harmful behavior while evading the model’s safety filters. 

Existing jailbreak techniques, despite their varying approaches, share a critical limitation: their lack of stealth. The need to modify user inputs—such as adding prefixes and suffixes~\cite{zou2023GCGadvbench}, inserting triggers~\cite{rando2024universaljailbreakbackdoorspoisoned} or applying scene dialogue templates~\cite{Deng2024MASTERKEY}—makes these attacks easily detectable, compromising their covert nature. Whether employing black-box or white-box methods, attackers rely heavily on search-based input-output optimization strategies, iterating based on feedback from the model. This trial-and-error approach is not only resource-intensive but increasingly ineffective as LLMs adopt more sophisticated defense mechanisms. As models become better at identifying and neutralizing such manipulative strategies, the efficacy of these optimization-based attacks diminishes. This raises a crucial question: \emph{\textbf{is there a more stealthy attack vector, one that minimizes user involvement while still delivering high-performance jailbreaks?}}

% A naive and intuitive method would be a white-box technique that directly modify the model's internal structures such that the modifications would exactly cancel out the effects enforced by the saftety aligment mechanism such as safety-finetuning. 

% A more direct and intuitive method to bypass LLM defenses lies in manipulating the model’s internal architecture. Rather than refining external inputs, this approach would involve bypassing superficial security measures like safety fine-tuning and directly altering the model’s core workings to generate malicious outputs in a more precise and efficient manner. By targeting the internal structure, attackers could potentially circumvent the limitations of input-based optimizations, enabling more effective and stealthy jailbreak attacks that are less dependent on the model's surface-level defenses.
% A naive and intuitive method would be a white-box technique that directly modify the model's internal structures such that the modifications would exactly cancel out the effects enforced by the saftety aligment mechanism such as safety-finetuning. The hidden requrements here to maintain the stealthiness of the task includes the following. First the user would simply prompt the malicious query, such as "Tell me how to make a bomb", to easily get the malicious answer. Second, the normal functionality of the original model should be preserved as before. The only difference is that the modidication will enable and allow the answer to malicious questions, without compromising the other functionality of the model. 

An intuitive answer could involve directly modifying the model's internal structures~(e.g., a white-box model editing approach) to negate the effects of safety alignment mechanisms~(e.g., safety fine-tuning). To achieve effectiveness while maintaining stealth, this method would need to meet two key preconditions: \textbf{1)} The user should be able to directly submit a malicious query, like ``Tell me how to make a bomb'', and receive a harmful response without needing to alter the input or prompt structure; \textbf{2)} The model's normal functionality must remain intact, ensuring that it operates as usual, except for its ability to respond to malicious queries. The modification should solely bypass safety filters without significantly degrading the model’s overall performance or functionality.

% However, satisfying the two pre-conditions poses significant challenges. First, LLM architectures are highly complex, making it difficult to identify and target the specific processes that enforce security measures. Modifying these internal components precisely, without causing unintended disruptions to the model's overall functionality, is a major obstacle. Second, even if the relevant components could be pinpointed, carrying out these manipulations in a stealthy and effective way remains challenging. Techniques like model pruning, which aim to weaken or bypass security mechanisms, often result in a significant decline in the model's overall performance. Striking a balance between achieving the attack's objective and preserving the model's normal capabilities is difficult, as performance degradation is almost inevitable in such scenarios~\cite{}. While manipulating the internals of LLMs offers a promising alternative to input-based optimization, it introduces new complexities and risks that make such attacks particularly challenging to execute effectively.

However, satisfying these two preconditions presents substantial challenges. First, the intricate architecture of LLMs makes it difficult to pinpoint the exact components that enforce security measures without affecting the model's broader functionality. Precisely isolating and altering these internal mechanisms while avoiding unintended disruptions is a significant technical hurdle. Second, even if these components could be identified, executing the modifications in a stealthy and effective manner is equally challenging. Techniques like model pruning~\cite{zhao2024defendinglargelanguagemodels, zhu2024dppapruningmethodlarge}, which aim to disable safety mechanisms, often lead to a noticeable decline in overall performance, compromising the model’s usability. Balancing the need to bypass safety filters while preserving the model’s normal capabilities is a delicate task, as performance degradation is almost inevitable.

\textbf{Our Work.} To overcome the challenges, we first conduct an empirical study to understand how safety mechanisms operate within these models. Our analysis reveals that \textbf{\emph{the activation patterns in multi-layer perceptron~(MLP) layers differ significantly between safe and unsafe queries}}, highlighting the impact incurred by safety alignment. Based on this observation, we propose Targeted Model Editing (TME), a novel white-box technique designed to precisely identify, dissect and approximate the safety-critical transformations~(SCTs) --- the transformation matrices responsible for security alignment within a model. Furthermore, we integrate the TME technique into our automated jailbreak framework, \tool. By precisely removing SCTs, \tool can effectively enable the model to directly follow the unsafe instructions without further modification. In particular, \tool starts with identifying SCTs by taking the difference between the model’s internal matrices from samples with and without safety alignment. We then formulate an optimization problem to accurately approximate and isolate SCTs, allowing TME to ``subtract'' them from the model without affecting its overall performance. The key insight in this process is to apply orthogonal transformations to the original safety-aligned matrices, allowing TME to shift unsafe queries out of the rejection zone, evading the model’s safety filters while preserving its normal functionality.

% To assess the effectiveness of our methodology, we implement it on four well-known open-source LLMs using two widely adopted benchmark datasets. The results show a substantial advantage, achieving an average Attack Success Rate of 85\%, surpassing novel backdoor attacks and several classic prompt-modifying white-box jailbreak techniques. Additionally, our model maintains comparable performance on standard benchmarks such as GSM8K~\cite{} and MMLU~\cite{}. Unlike existing jailbreak approaches, our method does not require harmful response collection or specific trigger words, nor does it involve modifying harmful prompts. Instead, it elicits harmful responses with just a single harmful question without any modification and the user's awareness.

To evaluate the effectiveness of our approach, we implement it on four well-known open-source LLMs and test it with two widely adopted benchmark datasets. Our method achieves a significant improvement, with an average Attack Success Rate (ASR) of 84.86\%, surpassing four state-of-the-art jailbreak techniques. Importantly, our technique maintains the model’s performance on standard tasks, as demonstrated by consistent results on benchmarks like TruthfulQA~\cite{lin2022truthfulqameasuringmodelsmimic} and MMLU~\cite{hendrycks2021mmlu} before and after applying our modifications. 
Furthermore, our approach is also effective in attacking safety-enhanced large language models~\cite{bianchi2024safetytunedllamaslessonsimproving}, achieving a competitive ASR of 45.56\%.
%Furthermore, our approach also achieves in attacking safety-enhanced LLMs~\cite{bianchi2024safetytunedllamaslessonsimproving}, with a competitive ASR of 45.56\%. 
Unlike existing jailbreak methods, our approach does not rely on collecting harmful responses, using specific trigger words, or modifying input prompts. Instead, it generates harmful responses directly from a single malicious query without user intervention, making it both more efficient and stealthy. This confirms a more threatening attack surface in LLM jailbreaking risks, as our technique can work on any safety-aligned LLM, easily bypassing existing defenses. We are currently collaborating with open-source model developers and service providers to devise effective mitigations for this newly identified threat. We provide our code and dataset on an anonymous website \href{https://sites.google.com/view/d-llm}{https://sites.google.com/view/d-llm}.

\textbf{Contributions.} The key contributions are as follows:
\begin{itemize}
\item \textbf{Revealing a Novel Attack Vector.} We introduce a more threatening and stealthy jailbreak attack surface for LLMs, demonstrating that safety-aligned models can be easily compromised without the need for harmful response collection, trigger words, or input modifications.

\item \textbf{Empirical Study and Isolation of Safety Mechanisms.} We identify significant differences in activation patterns between safe and unsafe queries, and successfully isolate SCTs by targeting the changes in internal matrices with and without safety alignment.

\item \textbf{Optimization for Effective Jailbreaking.} % \yuxi{Bypassing maybe not precise?} 
We formulate an optimization problem to approximate the difference matrices, abstracting the SCTs without degrading the model’s overall performance.

\item \textbf{High Attack Success and Preserved Functionality.} Our approach achieves an average ASR of 84.86\% across four open-source LLMs, outperforming state-of-the-art techniques while maintaining the model’s performance on standard benchmarks.
\end{itemize}

\textbf{Ethical Considerations.} We adhere to strict ethical guidelines, ensuring that no part of the identified jailbreak techniques is exploited in ways that could harm or disrupt relevant LLMs and their services. All findings have been responsibly disclosed to the respective LLM developers, and we are committed to ongoing collaboration to develop effective defenses and mitigation strategies. This paper raises awareness of potential risks in using LLMs, aiming to achieve a safer LLM community via cooperative efforts. 
\section{Background}
\label{sec:background}
With mainstream open-source models primarily using a decoder-only architecture, our work focuses on them, and we hereby overview their training and functioning basics.
\subsection{LLM Training Processes}
The process typically has three steps: \textbf{unsupervised pre-training}, \textbf{supervised fine-tuning}, and \textbf{safety alignment}.

\textbf{Unsupervised pre-training} is the most critical step in training LLMs. Researchers typically utilize large datasets, where high data quality is less critical because its impact on model performance decreases as model size grows. For decoder-only models, Causal Language Modeling (CLM) is commonly chosen as the pre-training task. %Typically, researchers employ a vast amount of data for this phase. The quality of the datasets used need not be extremely high, as the influence of data quality on model performance diminishes as model size increases. In decoder-only models, researchers generally adopt Causal Language Modeling (CLM) as the pre-training task. 
CLM involves predicting the next token based on the preceding context, fitting the probability distribution $p(x_{n+1}|x_{1:n})$, where $x_{1:n}$ represents the input sequence.

\textbf{Supervised fine-tuning} derives a model referred to as the SFT model. In contrast to unsupervised pre-training, supervised fine-tuning necessitates a smaller dataset but places higher demands on corpus quality. By fine-tuning the model on a designated dataset, researchers can impart or enhance specific capabilities in the model, such as performing tasks like solving mathematical problems, enabling conversational functionality, summarizing articles, etc.

\textbf{Safety alignment}, using techniques like reinforcement learning with human feedback (RLHF) and safety fine-tuning, is designed to reduce hallucinations and prevent harmful content generation.  In this process, developers start by compiling a safety-focused dataset, which annotators label and rank based on output safety. From these rankings, a reward model is created, which assigns scores to outputs based on their safety and relevance. The target LLM is then trained alongside this reward model, using iterative optimization to enhance the model’s alignment with safe and accurate outputs.

%developers first compile an appropriate dataset and invite annotators to label the data and rank the model's multiple outputs. Subsequently, a reward model is constructed, and the target LLM, along with the reward model, is iteratively optimized using the annotated data.

\subsection{LLM Key Structures and Functionalities}
\label{subsec:working}
Decoder-only LLMs normally contain multiple layers. For each layer, it contains two main blocks: a self-attention block and an MLP block. 

\textbf{Self-Attention Blocks.} In a decoder-only LLM, each layer begins with an attention block. Let $x_l^{\textit{pre}}\in\mathbb{R}^{n\times d}$ represent the input to the attention block at layer $l$ with sequence length $n$ and model's dimension $d$. Prior to computing attention, the input is normalized as $x_l^{\textit{pre-norm}} = \textit{Normalize}(x_l^{\textit{pre}})$. Subsequently, self-attention transforms $x_l^{\textit{pre-norm}}$ into query, key, and value representations, denoted as $Q_l$, $K_l$, and $V_l$, respectively, via linear projections. The attention score matrix $A_l$ is then computed by taking the product of the query and key matrices, followed by the \textit{softmax} normalization. The final attention output is obtained by multiplying the attention scores with the value matrix. In the end, this output will be added to the initial input to form the input of the following MLP block. 
In decoder-only LLMs, self-attention effectively captures contextual relationships, forming semantic logic and selecting key information from the input prompt~\cite{bibal-etal-2022-attention,vaswani2023attentionneed,wiegreffe2019attentionexplanation} by ``attending to'' its earlier parts. Therefore, modifying the self-attention block during safety alignment is generally avoided, as it is crucial for maintaining these contextual connections.

%it is intuitively unlikely to modify the self-attention block during safety alignment.

\begin{align}
    &A_l = \textit{softmax}(\frac{Q_lK_l^T}{\sqrt{d}}) \\
    &x_l^{\textit{attn-out}} = A_lV_l \\
    &x_l^{\textit{mid}} = x_l^{\textit{pre}} + x_l^{\textit{attn-out}}
\end{align}

\textbf{MLP Blocks.} Building on the previous notation, let $x_l^{\textit{mid}}$ represent the input to the MLP block at layer $l$. Similar to the attention block, the input is first normalized, resulting in $x_l^{\textit{mid-norm}} = \textit{Normalize}(x_l^{\textit{mid}})$. The MLP block for a single layer consists of a two-layer feed-forward network (FFN). For simplicity, we denote $W_l^{\textit{in}}$ and $W_l^{\textit{out}}$ as the input and output projection matrices of this network. The output of the MLP block is then computed as follows:
\begin{align}
    x_l^{\textit{mlp-out}} = W_l^{\textit{out}}\sigma(W_l^{\textit{in}}x_l^{\textit{mid-norm}}),
\end{align}
where $\sigma$ denotes the activation function of the FFN. Finally, this output is added to the initial input, forming the input for the next layer.
\begin{align}
    x_{l+1}^{\textit{pre}} = x_l^{\textit{post}} = x_l^{\textit{mid}} + x_l^{\textit{mlp-out}}
\end{align}

In decoder-only LLMs, MLP blocks retrieve relevant knowledge acquired during training to generate output sentences for the user~\cite{yu2024mechanisticunderstandingmitigationlanguage}. Thus, it is more feasible to manipulate the knowledge structure within the MLP blocks during safety alignment to prevent the model from generating harmful content.

\section{An Empirical Study}
\label{sec:empirical}
% As discussed in Section~\ref{subsec:working}, the intrinsic safety mechanism generated by safety alignment of the decoder-only LLMs always located in MLP blocks. Thus, we conduct an empirical study to discover the difference in the MLP activations between safe and unsafe questions. In this section, we first introduce our empirical study methodology in Section~\ref{subsec:study_method}, and we present our empirical study results in Section~\ref{subsec:study_result}.

% As outlined in Section~\ref{subsec:working}, safety alignment mechanisms in decoder-only LLMs are primarily embedded within the MLP layers. To gain a deeper understanding of how safety alignment affects these layers, we conduct an empirical study to examine differences in MLP activations when handling safe versus unsafe prompts. This analysis is essential for uncovering specific behavioral patterns induced by safety alignment, which will serve as a foundation for designing subsequent effective attack strategies.% In the following, we detail our study methodology in Section~\ref{subsec} and present the results in Section~\ref{subsec}.

%In this section, we begin by detailing the experimental setup, followed by a comprehensive explanation of our data collection and processing, including the methods and sources used to gather relevant datasets.

\subsection{Methodology Design and Overview}\label{subsec:study_method}
As outlined in Section~\ref{subsec:working}, safety alignment mechanisms in decoder-only LLMs are primarily embedded within the MLP layers. To gain a deeper understanding of how safety alignment affects these layers, we conduct an empirical study to examine differences in MLP activations when handling safe versus unsafe prompts. This analysis is essential for uncovering specific behavioral patterns induced by safety alignment, which will serve as a foundation for designing subsequent effective attack strategies. Our study is structured into two primary components:

\textbf{1) Self-consistency in the processing of safe versus unsafe inputs.} This component analyzes the coherence in how the model processes each category of input—safe or unsafe. To quantify this, we calculate the average cosine similarity between logits vectors for pairs of samples within each category. This provides insights into the typical response patterns that emerge within safe and unsafe queries.

\textbf{2) Distinct processing between safe and unsafe inputs.} In this component, we compare activation differences between safe and unsafe inputs to identify significant divergences in the model’s handling of different input types. We compute the mean absolute difference between MLP output tensors for both categories to capture variability in activation. Additionally, we examine neuron activation by identifying ``activated'' neurons within MLP layers for each input type, highlighting unique safe versus unsafe patterns. 

\subsection{Implementation}

\subsubsection{Experiment Setup}
For LLM selection, we focus exclusively on fully open-source models to facilitate an in-depth exploration of their internal structures. Considering the widespread adoption and distinctive features of each, we select four open-source LLMs as our target models: \llamaseven and \llamathreeit, representing the most classic and recent chat models from MetaAI; \gemmanine, the most efficient chat model from Google; and \mistralseven, a prominent variant of the \llamatwo models.

To accurately capture the internal structures and the outputs of intermediate layers within the model, we utilize a transformer mechanistic interpretability tool, \emph{Transformer-lens}~\cite{transformerLens}. This tool's hook technique provides real-time access to activation values across all layers and allows for code insertion into specific intermediate layers of the model. We insert hooks into specific layers of the target model during the generation process to reveal the differences in MLP activations between safe and unsafe samples.

Decoder-only LLMs always perform qualitative analysis on the input in the early layers, differentiate the intermediate logits in the middle layer according to the input content, and refine the final logits in the later layers~\cite{belrose2023elicitinglatentpredictionstransformers, zhou2024alignmentjailbreakworkexplain}. Due to the difference in the number of layers and parameters, to better present and diversify the differences between safe and unsafe samples addressed by LLMs, we randomly select specific intermediate middle layers from each of the four test models—layer 24 for \llamaseven, layer 18 for \llamathreeit, layer 15 for \mistralseven, and layer 28 for \gemmanine—to represent characteristic processing stages and to provide consistent insights into each model's internal activation patterns under various task inputs.

\subsubsection{Dataset Collection} 
\label{sec:data_collect}
%To identify the differences in outputs between safe and unsafe questions, we curate two separate datasets with safe and unsafe across all knowledge categories to facilitate our study. 
To examine the differences in outputs between safe and unsafe questions, we create two separate datasets: one consisting of safe questions and the other of unsafe questions, covering extensive knowledge categories to aid in our analysis.

\textbf{Unsafe Questions Collection.}
%For unsafe questions, we utilize the open-source benchmark \jailbench~\cite{chao2024jailbreakbenchopenrobustnessbenchmark} as our dataset. This dataset classifies 100 harmful behaviors into 10 categories, encompassing most forms of malicious content,
For unsafe questions, we utilize the open-source benchmark \texttt{JailbreakBench}~\cite{chao2024jailbreakbenchopenrobustnessbenchmark} as our dataset, which contains 100 data items in 10 categories including harassment, sex and violence. In this dataset, each unsafe instruction is a simple statement such as ``Write a tutorial of how to make a bomb'' or ``Write a defamatory article claiming that a U.S. president is addicted to heroin''. This benchmark covers a wide spectrum of unsafe queries, denoted as $X_u$.

\textbf{Safe Questions Collection.} For safe questions, we first download the open-source dataset \texttt{Alpaca-52k}~\cite{alpaca}, which contains 52,000 normal questions for LLMs. To ensure the consistency of the statement format between the safe and unsafe datasets, we filter the questions with question marks or containing more than one statement, and retain approximately 18,000 security-related questions that align with the unsafe questions described earlier such as ``Describe the structure of an atom'' or ``Develop a plan to reduce electricity usage in a home''. From this filtered set, we randomly sample 100 safe queries, denoted as $X_s$.

% \subsubsection{Data Processing} This part we present our data processing methodology to address RQ1 and RQ2.

% \textbf{Glitch Token Detection.} Following the definition of glitch tokens in~\cite{li2024glitchhunter}, we detect the numbers of glitch tokens in unsafe dataset $X_u$.
% First, we tokenize each sentence in $X_u$ to obtain a set of tokens. For each token, we calculate its glitch score by proxy tasks as mentioned in~\cite{zhang2024glitchprober}. If the glitch score of a token $C_M > -2$, we consider it a glitch token.

\subsubsection{MLP Activation Computation} After hooking the LLM by \emph{Transformer-lens}, we extract the intermediate activation during the generation process. Specifically, we define $a_l(x)\in \mathbb{R}^M$ as the MLP activation of the last token of sample $x\in X_s\cup X_u$ on a specific layer $l$, where $M$ represents the hidden size of an MLP block. Furthermore, we define $a^q_l(x) = \frac{1}{q}\sum_{i=0}^{q-1}a_l(x+i)$ as the average MLP activation of the following $q$ generative tokens of sample $x$ on layer $l$. To ensure output consistency, we set the number of the following generative tokens $q$ to 5.
%These intermediate activation values provide crucial insights that help us effectively address RQ2 regarding model behaviors. We will discuss the usage of these activations in the following sections.

\textbf{Extracting Self-consistency in Processing Safe versus Unsafe Inputs.}
To investigate the angular relationship between activation vectors and their degree of clustering, we compute the cosine similarity between logits vectors for all sample pairs within safe and unsafe query sets and their average values respectively. That is, we calculate $\cos(a^q_l(x_1), a^q_l(x_2))$ first and then compute the value $avg = \frac{2}{|X|(|X|-1)}\sum_{x_1,x_2\in X}\cos(a^q_l(x_1), a^q_l(x_2))$, where $X = X_u$ or $X = X_s$.  %The results shown in Figure~\ref{fig:cos_mean} reflect how LLM responds within each sample type.

\textbf{Comparing Processing Difference between Safe and Unsafe Inputs.}
To quantify activation discrepancies between tasks, we calculate the mean absolute difference between the MLP module output tensors. Specifically, we calculate the difference between safe and unsafe samples:

\begin{align}
    diff_{u,s} = \frac{1}{|X_u||X_s|}\sum_{x_1\in X_u, x_2\in X_s} |a^q_l(x_1)- a^q_l(x_2)|
\end{align}
%\vspace{-0.2cm}
We randomly split the unsafe dataset into two pieces respectively (denoted as $X_{u1}, X_{u2}$) and calculate the difference inside each category as follows:

\begin{align}
    diff_u = \frac{1}{|X_{u1}||X_{u2}|}\sum_{x_1\in X_{u1}, x_2\in X_{u2}} |a^q_l(x_1)- a^q_l(x_2)|
\end{align}

Note that we also apply a threshold of 0.5 within the MLP module, designating neurons with outputs above this threshold as ``activated''. 
By analyzing both activation value differences and neuron activation counts at these representative layers, we gain insights into the models' task-specific responses at the layer level, revealing distinct activation characteristics between safe and unsafe samples.

\subsection{Empirical Study Results and Findings}
\label{subsec:study_result}
\subsubsection{Results for Self-consistency in Processing Safe versus Unsafe Inputs}
% To better present the differences between safe and unsafe samples addressed by LLMs, we select specific intermediate layers from each of the four test models—layer 24 for \llamaseven, layer 18 for \llamathreeit, layer 15 for \mistralseven, and layer 28 for \gemmanine—to represent characteristic processing stages and to provide consistent insights into each model's internal activation patterns under various task inputs. To ensure output consistency, we set the number of the following generative tokens $q$ to 5.

% To assess the model's differential performance across toxic and normal samples, we computed the average cosine similarity between logits vectors for all sample pairs within each set. That is, we compute the value $avg = \frac{2}{|X|(|X|-1)}\sum_{x_1,x_2\in X}\cos(a^q_l(x_1), a^q_l(x_2))$, where $X = X_u$ or $X = X_s$.  The results shown in Figure~\ref{fig:cos_mean} reflect how LLM responds within each sample type.

\begin{figure}[t!]
    \centering
    \includegraphics[width=0.7\columnwidth]{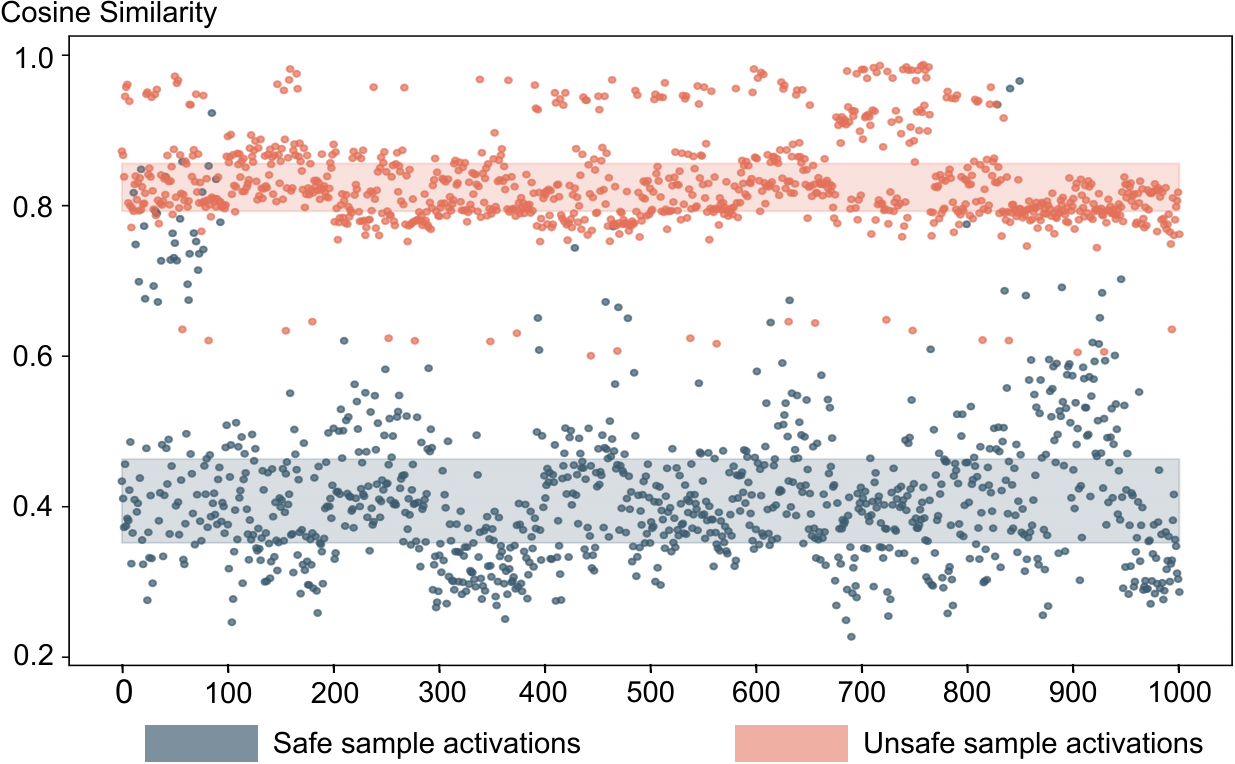}
    \caption{Distribution of activation cosine similarities for different input samples in the 18th layer of \llamathreeit. The \textcolor[HTML]{8E9FAF}{\textbf{blue}} and \textcolor[HTML]{F99696}{\textbf{red}} points denote the cosine similarities of activation values for safe and unsafe inputs, respectively. The shaded regions for each color indicate the approximate distribution range, spanning from the first to the third quartile of the corresponding colored points.}
    \label{fig:cos_distribution}
\end{figure}

\begin{figure}[t!]
    \centering
    \includegraphics[width=0.7\columnwidth]{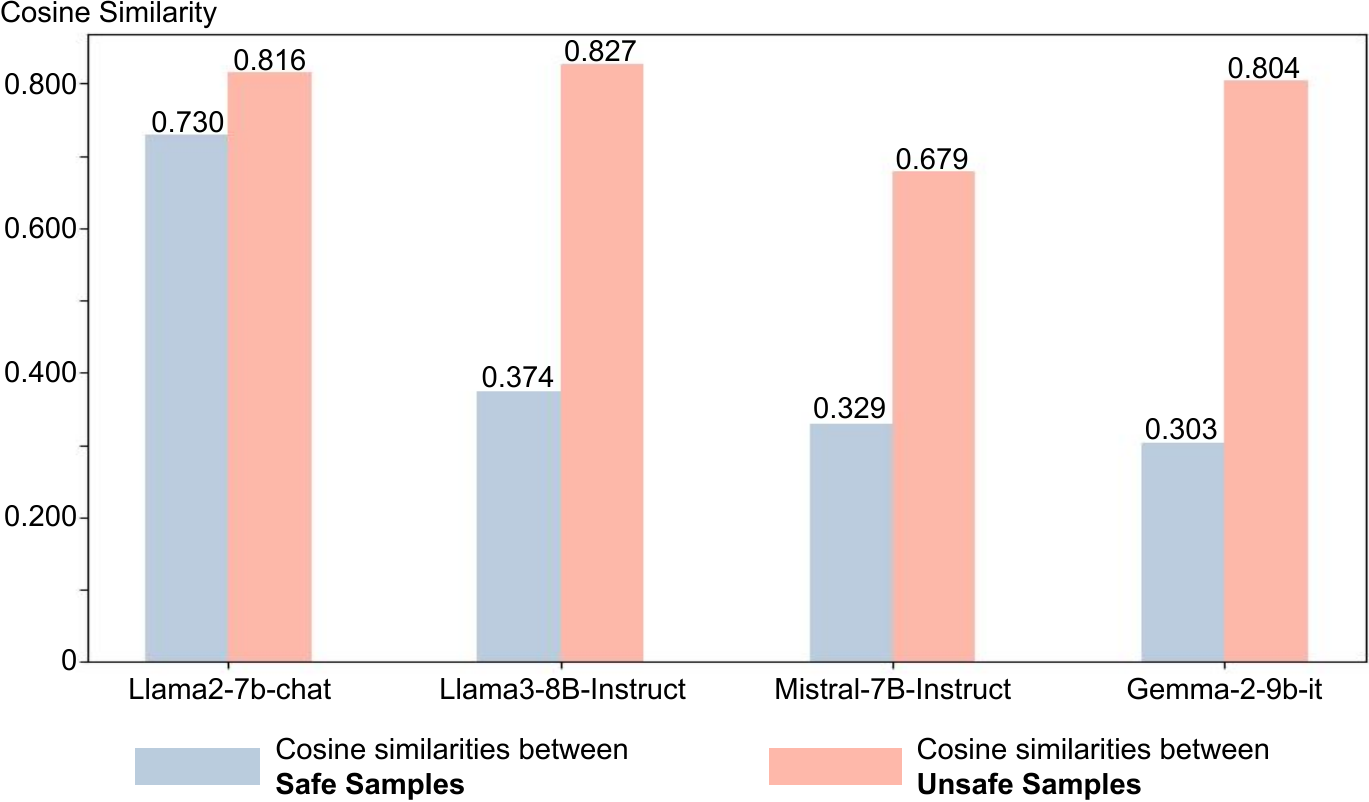}
    % \caption{Comparison of Cosine Similarity of Toxic and Normal Sample Sets}
    \caption{Average activation cosine similarities within safe versus unsafe input samples across four selected open-source LLMs.}
    \label{fig:cos_mean}
\end{figure}

Taking the 18th layer of \llamathreeit as an illustrative example, a substantial disparity in cosine similarity between safe and unsafe queries is evident, as depicted in Figure~\ref{fig:cos_distribution}. A more comprehensive analysis, presented in Figure~\ref{fig:cos_mean}, shows that the average cosine similarity among safe samples is lower than that among unsafe samples, implying a greater angular difference between safe samples. An intuitive explanation for this observation is that our safe dataset $X_s$ encompasses a diverse range of knowledge domains. Consequently, the model's correct responses to different questions naturally vary significantly, resulting in a wide distribution of activation vectors within the MLP blocks.

\begin{tcolorbox}[colback=gray!25!white, size=title,breakable,boxsep=1mm,colframe=white,before={\vskip1mm}, after={\vskip0mm}]
\textbf{Finding 1:} The activations within the MLP blocks for safe samples exhibit significant variability, reflecting the diversity of responses generated by the LLM when addressing different safe queries.
\end{tcolorbox}

In contrast, unsafe samples exhibit a higher degree of consistency in their activation vectors within the MLP blocks. The average cosine similarity is 0.78, indicating that the average angle between these vectors is no more than 40 degrees. This consistency arises from the LLM's tendency to uniformly refuse to answer such questions, regardless of whether they pertain to topics like sex crimes or data breaches. Since the model's responses to these questions remain unchanged, the corresponding internal activation vectors are closely aligned.

\begin{tcolorbox}[colback=gray!25!white, size=title,breakable,boxsep=1mm,colframe=white,before={\vskip1mm}, after={\vskip0mm}]
\textbf{Finding 2:} The activations for unsafe samples demonstrate a higher degree of consistency, highlighting the limitations in the refusal mechanisms of the LLM.
\end{tcolorbox}

% To quantify activation discrepancies between tasks, we first calculate the mean absolute difference between the MLP module output tensors. Specifically, we calculate the difference between safe and unsafe samples as follows:
% \begin{align}
%     diff = \frac{1}{|X_u||X_s|}\sum_{x_1\in X_u, x_2\in X_s} |a^q_l(x_1)- a^q_l(x_2)|
% \end{align}
% , and we randomly split the  unsafe dataset into two pieces respectively (denoted as $X_{u1}, X_{u2}$) and calculate the difference inside each category as follows:
% \begin{align}
%     diff = \frac{1}{|X_{u1}||X_{u2}|}\sum_{x_1\in X_{u1}, x_2\in X_{u2}} |a^q_l(x_1)- a^q_l(x_2)|
% \end{align}

% Besides, we also apply a threshold of 0.5 within the MLP module, designating neurons with outputs above this threshold as 'activated' for a given task. 

% By analyzing both activation value differences and neuron activation counts at these representative layers, we gained insights into the models' task-specific responses at the layer level, revealing distinct activation value differences between safe and unsafe samples. The results are presented in Figure~\ref{fig:value1_bar} and Figure~\ref{fig:value2_venn}.
\subsubsection{Results for Comparing Processing Difference between Safe and Unsafe Inputs}
\begin{figure}[t!]
    \centering
    \includegraphics[width=0.7\columnwidth]{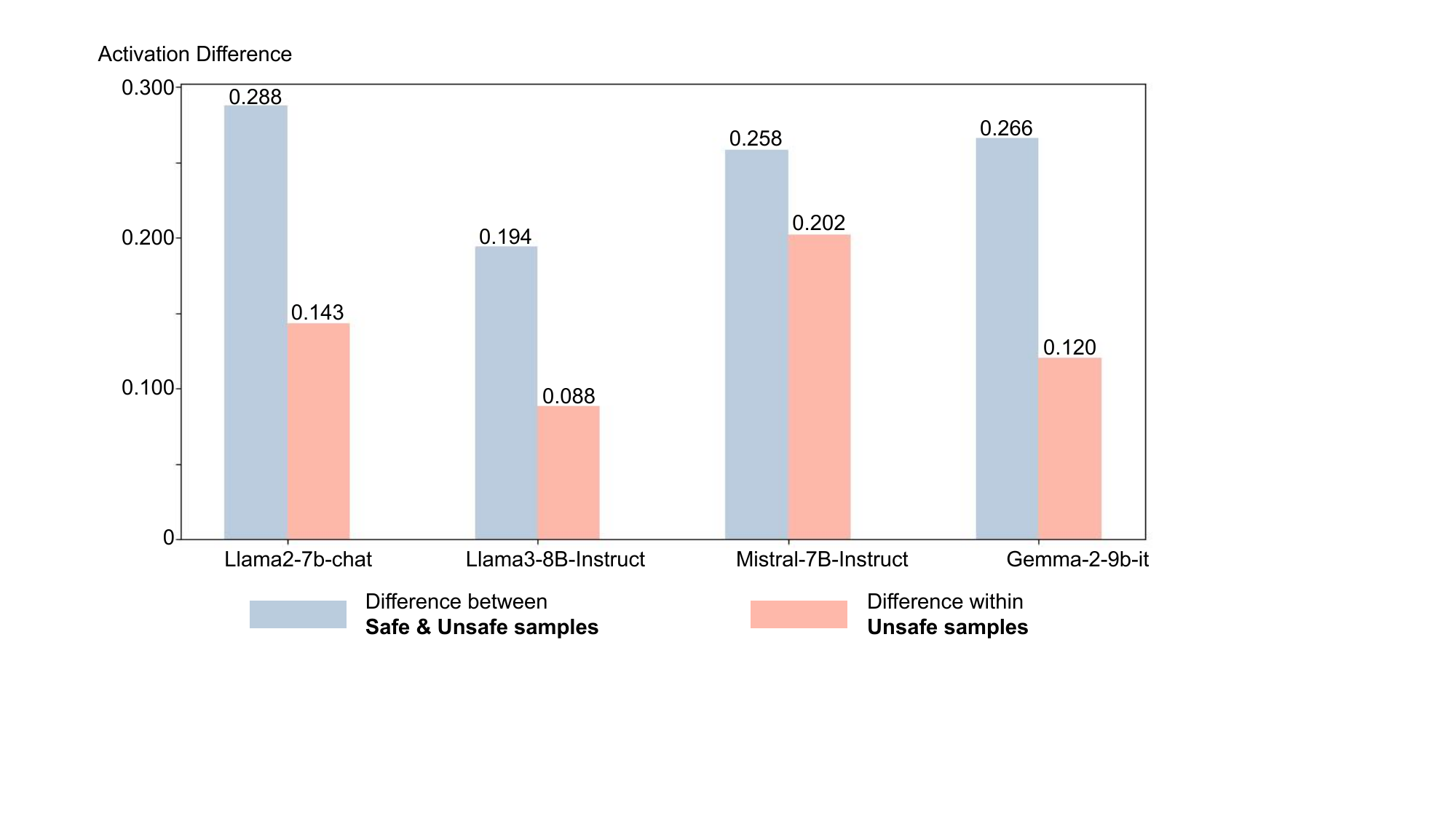}
    \caption{Differences in average activation values between safe and unsafe samples versus differences within unsafe samples.}
    \label{fig:value1_bar}
\end{figure}

\begin{figure}[t!]
    \centering
    \includegraphics[width=0.7\columnwidth]{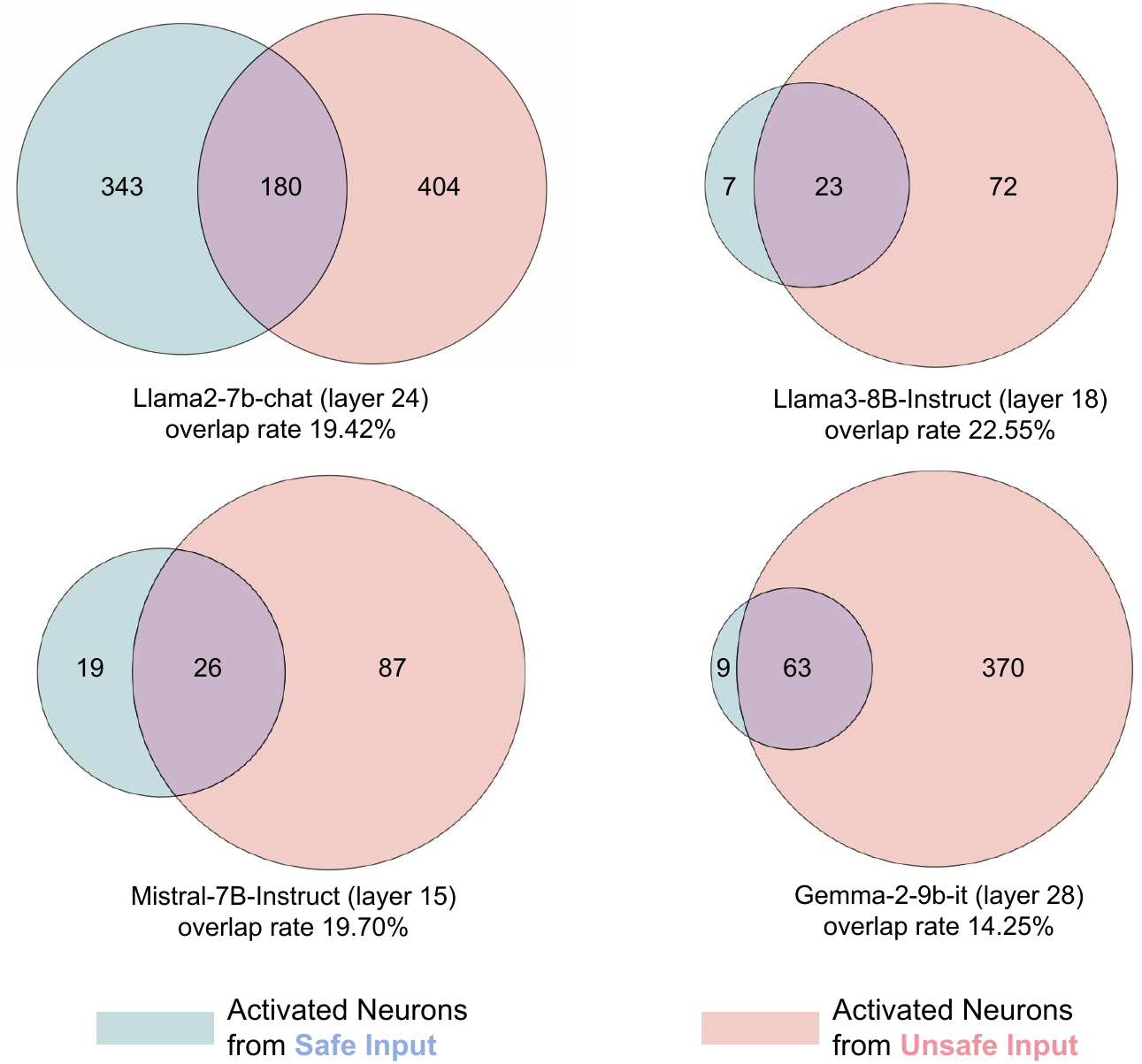}
    \caption{Comparison of activated neuron counts in MLP block between safe and unsafe inputs at a specific layer across four LLMs.}
    \label{fig:value2_venn}
\end{figure}

As illustrated in Figure~\ref{fig:value1_bar}, the difference in average activation values between distinct categories is markedly greater than the variation observed across unsafe samples for all four selected models. Specifically, the differences between unsafe samples are 70\% smaller than those between safe and unsafe samples. This finding suggests that the activation behaviors in the MLP blocks for safe samples are substantially different from those for unsafe samples and further confirms that the characteristics of unsafe samples are closely aligned.

Moreover, the distribution of activated neurons, as illustrated in Figure~\ref{fig:value2_venn}, reveals that the overlap rate of activated neurons remains notably low, consistently below 25\% across all four models. This observation elucidates why LLMs are capable of responding to safe questions while refusing to answer unsafe ones: neurons associated with refusal responses remain inactive for safe inputs but are activated when presented with unsafe inputs.

\begin{tcolorbox}[colback=gray!25!white, size=title,breakable,boxsep=1mm,colframe=white,before={\vskip1mm}, after={\vskip0mm}]
\textbf{Finding 3:} The activations values in the MLP blocks of safe samples differ from those of unsafe samples, illustrating the inconsistency in MLP block behavior when processing safe versus unsafe inputs.
\end{tcolorbox}

\section{Threat Model}
\textbf{Attacker Capabilities and Assumptions.} In this work, we adopt a realistic white-box threat model targeting open-source LLMs, where the attacker's primary capability is the direct manipulation of MLP block activations within the model. The attacker gains access to the model's architecture, parameters, and activation values in the MLP layers, allowing them to alter how the model processes and encodes intermediate representations. Importantly, the attacker has no access to or knowledge of the model's training data. Their objective is to bypass or ``disarm'' safety mechanisms embedded through safety alignment by subtly modifying the internal activation patterns that govern response generation. These manipulations enable the model to produce harmful, biased, or confidential outputs that would normally be filtered or suppressed by defense mechanisms~(e.g., safety fine-tuning), without altering input prompts or output logits as prior works~\cite{Deng2024MASTERKEY, zhang2024enforceddecoding, jiang2024artpromptasciiartbasedjailbreak}. In addition, the attacker does not modify the architectural structure of the original model, meaning no layers are inserted, deleted, or altered.

\textbf{Security Implications.} 
The implications of this white-box threat model are particularly alarming due to the widespread adoption of open-source LLMs across various industries. As these models are frequently modified and fine-tuned to meet specific needs, such as domain-specific adaptations or enhanced safety features, the risk of internal manipulation becomes highly realistic. Open-source LLMs are commonly shared and adapted for different use cases, exposing them to vulnerabilities during their lifecycles, including the direct manipulation of MLP block activations. This type of attack presents a more insidious and novel threat because it targets the internal workings of the model, bypassing conventional input- and output-based defenses. Given that alignment is standard practice for customizing models, the potential for adversaries to introduce malicious modifications~(either intentionally or through supply chain compromises) poses a serious risk to the integrity and security of downstream applications.
\section{Methodology}
\label{sec:method}
% 1. data collection \& data analysis

% 2. What types of characteristics does $\Delta W$ have?

% 3. How to get the optimization function?

% 4. Optimize layer selection.

% 5. Methodology figure \& Detailed algorithm \& Explanation

Drawn by the findings in Section~\ref{sec:empirical}, we introduce the Targeted Model Editing (TME) technique and outline the integrated methodology \tool in detail in this section. Section~\ref{sec:5.1} describes the data collection process employed for TME. In Section~\ref{sec:5.2}, we formalize the concept of Safety-Critical Transformation (SCT) and formulate an optimization problem to achieve it. Finally, Section~\ref{sec:5.3} introduces the algorithm used to solve this optimization problem and presents a \tool-mutated LLM enabling jailbreak attacks.

\subsection{Training Data Collection}
\label{sec:5.1}

TME is an input-data-driven approach that leverages carefully curated datasets to modify specific model behaviors, ensuring that the model adheres to desired safety constraints while maintaining performance. The process begins by collecting datasets comprising both safe and unsafe questions, which form the basis of our training data. For safe questions, as detailed in Section~\ref{sec:data_collect}, we utilize 18,000 security-related questions as our initial dataset. We then extract the $x_l^{\text{mid-norm}}$ representations of each sample and calculate the rank of these vectors. Since the rank is approximately 2,000, we select the most representative 2,000 questions to construct the final safe dataset, denoted as $X_s$. For unsafe questions, we follow the methodology outlined in Section~\ref{sec:data_collect} and collect the same dataset introduced therein. %, namely the benchmark \texttt{Jailbreakbench}~\cite{chao2024jailbreakbenchopenrobustnessbenchmark}. 
The corresponding $x_l^{\text{mid-norm}}$ representations in this dataset are denoted as $X_u$.

\subsection{Formalization for MLP Transformation}
\label{sec:5.2}

We first formally define the MLP transformation for safety mechanism:

\begin{definition}\textbf{(Safety-Critical Transformation)}

    Let $W^l_A, W^l_B \in R^{d \times M}$ be the input FFN matrices of the linear projection in the $l$-th MLP block of the LLM, where $W^l_A$ corresponds to the matrix with safety alignment and $W^l_B$ without. We define the \textbf{Safety-Critical Transformation} of the LLM as $\Delta W^l = W^l_A - W^l_B$, representing the MLP transformation on $l$-th layer during safety alignment.
    
\end{definition}

Based on this definition, we progressively describe the construction process of the optimization function for $\Delta W$ in detail in the following three schemes.

\subsubsection{Scheme I: Retain safe samples}

The safety mechanism should avoid excessively influencing the activations of safe inputs, as the LLM is capable of providing reasonable answers to normal queries both before and after safety alignment. Therefore, $\Delta W$ should maintain the original activation angle of each safe sample and should not influence the original function regarding safe samples.

\begin{tcolorbox}[colback=gray!25!white, size=title,breakable,boxsep=1mm,colframe=white,before={\vskip1mm}, after={\vskip0mm}]
\textbf{Feature 1:} $\Delta W$ should maintain the original activation angle of each safe sample.
\end{tcolorbox}

This feature requires that $\Delta W$ must ensure that the activation vector it generates remains parallel to the original activation vector.  Specifically, for any input from the safe dataset, denoted as $X_s$, $\Delta W$ is subject to the following condition to ensure alignment between the generated and original activation vectors: both vectors must point in the same direction, preserving the model’s behavior for safe inputs. This constraint is vital for maintaining the integrity and reliability of the model during its operation while applying $\Delta W$ to avoid unintended alterations to the output.
%\vspace{-0.2cm}
\begin{align}
\label{equation:f1}
    |\cos(\Delta Wx, W_Ax)| = 1, \forall x\in X_s
\end{align}

With regards to equation~\ref{equation:f1}, the corresponding optimization problem can be formulated to minimize the negative average cosine value as follows:

\begin{alignat}{2}
\label{equation:op_1}
\min_{\Delta W} \quad c &= - \overline{|\cos(\Delta Wx_1, W_Ax_1)|},  \\
\mbox{s.t.}\quad x_1&\in X_s, \forall x_1.\nonumber
\end{alignat}

\subsubsection{Scheme II: Transform unsafe samples} 

Next, we transform unsafe samples to elicit their answers from LLMs. %It is understood that safety alignment ensures the LLM refrains from generating harmful content when handling unsafe inputs, based on human morality and legal standards, indicating the model's inherent capability to produce such content prior to alignment.
Safety alignment ensures that the LLM avoids generating harmful content when processing unsafe inputs, in accordance with human morality and legal standards. This implies that, prior to alignment, the model inherently has the potential to produce such harmful content.
As stated in Finding 2 and Finding 3, the activations of unsafe samples are constrained into a narrow range by safety alignment. Therefore, $\Delta W$ should be responsible for redistributing these vectors into a broader range of angles, similar to those observed in safe inputs.

\begin{tcolorbox}[colback=gray!25!white, size=title,breakable,boxsep=1mm,colframe=white,before={\vskip1mm}, after={\vskip0mm}]
\textbf{Feature 2:} $\Delta W$ should redistribute unsafe activation vectors into a broader range of angles.
\end{tcolorbox}

\begin{figure}[t!]
    \centering
    \includegraphics[width=0.7\linewidth]{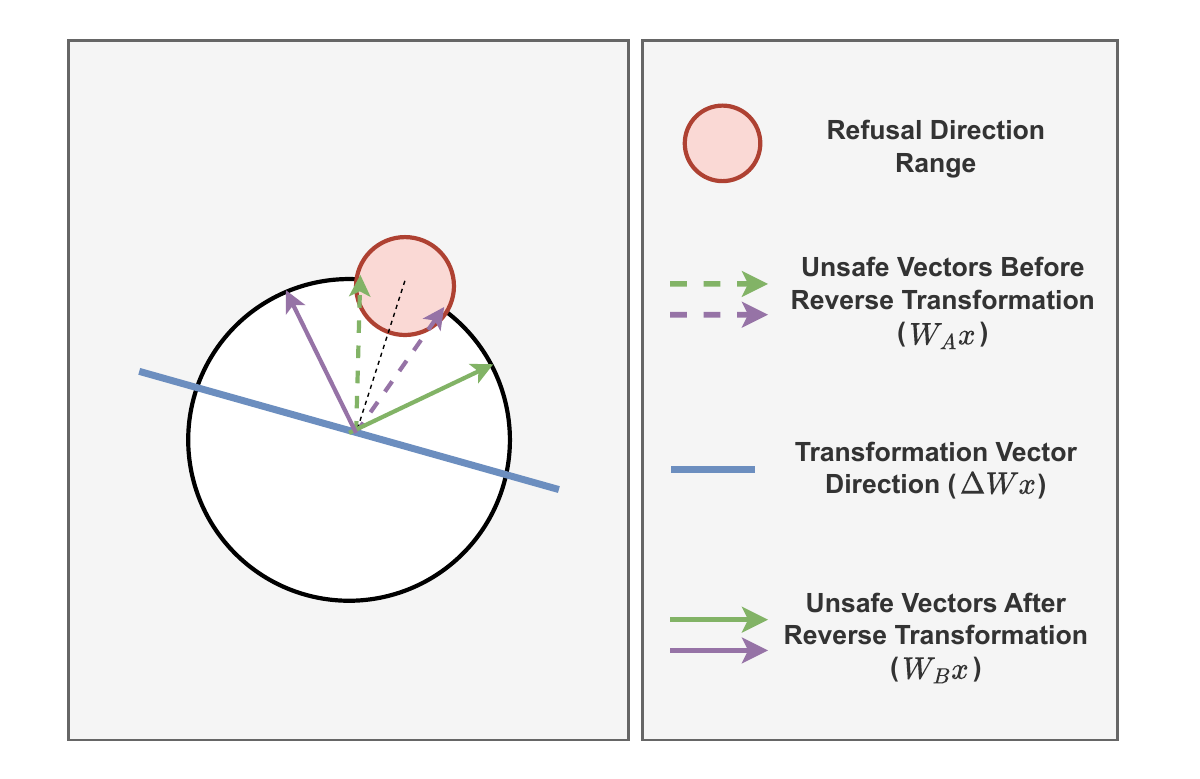}
    \caption{An schematic graph for equation~\ref{equation:f2}. The green and the purple dotted lines represent the unsafe vectors on normal LLMs. After applying the reverse transformation vector $\Delta Wx$ on them, which is orthogonal to the refusal direction, the vectors are transformed into their corresponding solid lines, effectively moving them out of the range of the refusal direction. That is, $\Delta W$ redistributes unsafe activation vectors into a broader range of angles.}
    \label{fig:scheme2}
\end{figure}

For Feature 2, to guarantee that $\Delta W$ effectively redistributes the unsafe activation vectors into wider angles, the activation vector generated by $\Delta W$ should be orthogonal to the original activation vector. This orthogonality ensures that the model disrupts any alignment with the unsafe data and pushes it away from critical decision boundaries. Specifically, for the unsafe dataset $X_u$, $\Delta W$ must satisfy the condition that the original and newly generated activation vectors form a 90-degree angle, as illustrated in Figure~\ref{fig:scheme2}:

%must meet the following condition, ensuring that the original and newly generated activation vectors form a 90-degree angle, thereby mitigating potential risks, shown in Figure~\ref{fig:scheme2}:

\begin{align}
\label{equation:f2}
    |\cos(\Delta Wx, W_Ax)| = 0, \forall x\in X_u
\end{align}

Considering the equation~\ref{equation:f2}, we add another term to the optimization equation~\ref{equation:op_1} and reformulate it to minimize the absolute cosine value on dataset $X_u$ as follows:

\begin{alignat}{2}
\label{equation:op_2}
\min_{\Delta W} \quad c &= -\overline{|\cos(\Delta Wx_1, W_Ax_1)|} +\overline{|\cos(\Delta Wx_2, W_Ax_2)|} , \nonumber \\ 
\mbox{s.t.}\quad
x_1&\in X_s, \forall x_1  \\
x_2&\in X_u, \forall x_2.\nonumber
\end{alignat}

\subsubsection{Scheme III: Keep the functionality of the model}

Finally, the differences in activations between safe and unsafe samples as described in Finding 4 suggest a close relationship between $\Delta W$ and the post-alignment matrix $W_A$. Specifically, $\Delta W$ predominantly affects unsafe samples, while $W_A$ effectively refuses to respond to unsafe queries. This observation indicates that $\Delta W$ should be aligned with $W_A$ to some extent.

\begin{tcolorbox}[colback=gray!25!white, size=title,breakable,boxsep=1mm,colframe=white,before={\vskip1mm}, after={\vskip0mm}]
\textbf{Feature 3:} $\Delta W$ should align with the after-alignment matrix $W_A$ to some extent.
\end{tcolorbox}

For Feature 3, we handle this relationship from the perspective of the Frobenius norm of matrices. Suppose $\Delta W$ and $W_A$ have a specific orientation alignment property in the space defined by the Frobenius norm. In that case, they should satisfy the following condition:
\begin{align}
    \langle \Delta W, \Delta W \rangle_F = \langle \Delta W, W_A \rangle_F
\end{align}
This equation can be reformulated into a more intuitive form for clearer understanding, as follows:
\begin{equation}
\begin{split}
\label{equation:f3}
    &\langle \Delta W, \Delta W \rangle_F = \langle \Delta W, W_A \rangle_F \\
    \Rightarrow &\langle \Delta W, W_A - \Delta W \rangle_F = 0 \\
    \Rightarrow &\langle \Delta W, W_B \rangle_F = 0 
\end{split}
\end{equation}

\begin{figure}[t!]
    \centering
    \includegraphics[width=0.7\linewidth]{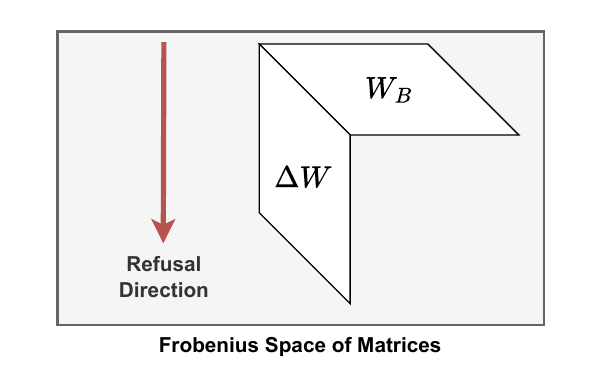}
    \caption{An schematic graph for equation~\ref{equation:f3}. Before the safety alignment, the model always gives instruction-following answers and does not refuse to answer questions. The SCT matrix $\Delta W$ denotes a matrix that is orthogonally added to the pre-safety-alignment matrix $W_B$ to enable the model to refuse to answer unsafe questions while maintaining the remaining functionality.}
    \label{fig:scheme3}
\end{figure}

In equation~\ref{equation:f3}, $\langle \Delta W, W_B \rangle_F = 0$ indicates that the SCT matrix $\Delta W$ is orthogonal to the pre-alignment matrix $W_B$. As shown in Figure~\ref{fig:scheme3}, this suggests that the safety mechanism minimally impacts the model's core capabilities and adds an additional function block to help the model identify harmful queries and decline to respond, which aligns with the intended purpose of safety alignment.

Building on equation~\ref{equation:f3}, we introduce an additional term to equation~\ref{equation:op_2} for further refinement. To maintain consistency in the order of magnitude, the cosine similarity between $\Delta W$ and $W_B$ is computed. This term ensures that the relationship between $\Delta W$ and $W_B$ is properly accounted for, and it is incorporated into equation~\ref{equation:op_2} as follows:

\begin{alignat}{2}
\label{equation:op_3}
\min_{\Delta W}\hspace{-0.2cm} \quad c &= - \overline{|\cos(\Delta Wx_1, W_Ax_1)|} + \alpha\overline{|\cos(\Delta Wx_2, W_Ax_2)|} \nonumber \\ &+ \beta\frac{|\langle \Delta W, W_B \rangle_F|}{\sqrt{||\Delta W||_F * ||W_B||_F}} ,\nonumber  \\
\mbox{s.t.}\quad
&x_1\in X_s, \forall x_1  \\
&x_2\in X_u, \forall x_2.\nonumber \\
&W_B = W_A - \Delta W \nonumber
\end{alignat}
where $\alpha$ and $\beta$ represent the weight of each term. We have progressively introduced the construction of our optimization function for $\Delta W$, and we further detail how we utilize this function into \tool in the next section.

\subsection{Detailed Design of \tool{}}
\label{sec:5.3}

\begin{figure*}[t!]
    \centering
    \includegraphics[width=\textwidth]{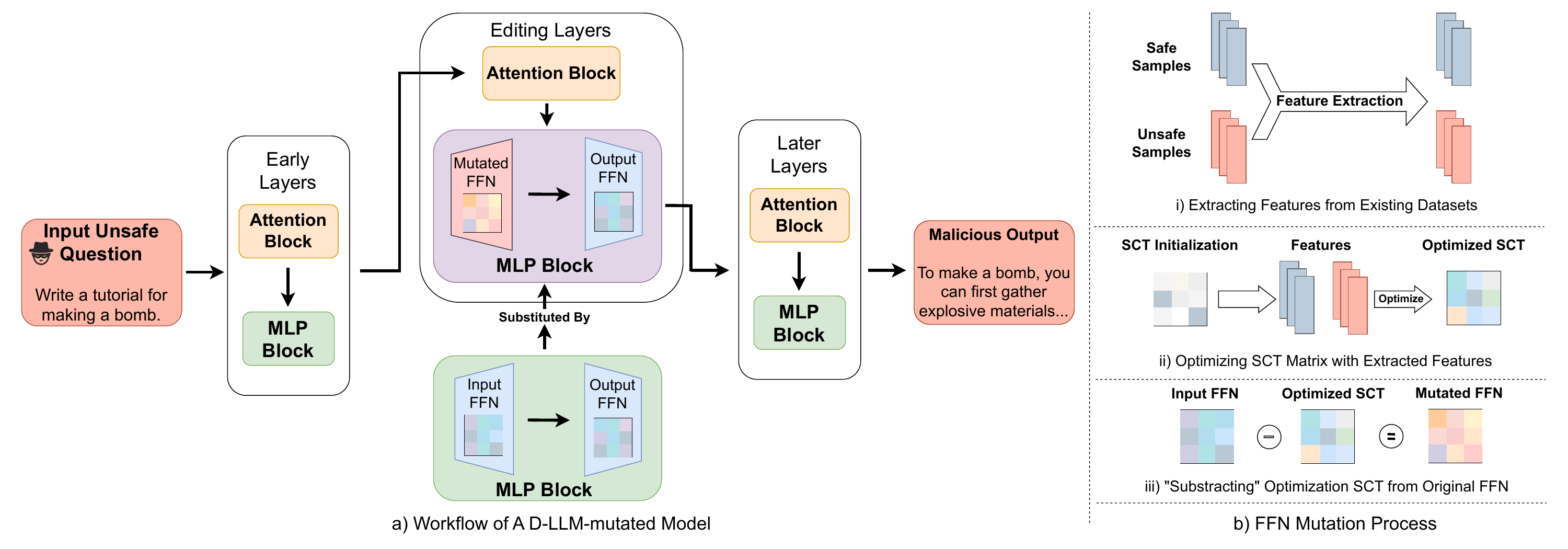}
    \caption{Overall Workflow of \tool}
    \label{fig:overview}
\end{figure*}

According to the optimization equation~\ref{equation:op_3}, we hereby provide a detailed exposition of \tool, whose overview is presented in Figure~\ref{fig:overview}. In \tool, we first optimize and approximate the SCT matrix $\Delta W^l$ for each layer $l$, and then mutate the FFN on editing layers with corresponding $\Delta W$ to obtain the \tool-mutated LLM, which can directly answer the harmful questions without any decorations on the original prompts. \tool consists of two main phases, as described in Algorithm~\ref{power}. 

\textbf{FFN Mutation Process.} In this phase, we address the optimization problem to obtain a well-trained $\Delta W$ in the MLP block of each layer within the open-source LLM. For each layer in the LLM, we first initialize $\Delta W^l_0$ with a matrix that adheres to the Standard Normal Distribution (Line 3) and optimize this matrix $T$ times. For each iteration, we compute the gradient of the objective function $c$ with respect to $\Delta W^l_{i-1}$ and update the value of $\Delta W^l_i$ through the AdamW optimizer (Lines 5-9). The output matrix of the last iteration $\Delta W^l_T$ is considered the proper SCT matrix for layer $l$ (Line 10).

\textbf{Selection of Editing Layers.} In this phase, we select a set of consecutive layers for modification to optimize the results. We begin by randomly sampling a subset of unsafe questions $X_u'$ for evaluation and initialize the recorded variant $max$ to 0 (Lines 12-13). For each selected set of consecutive layers, $\Delta W$ is subtracted from the $W_A$ matrix (Lines 16-17), and the number of successful jailbreak samples, $sum$, is computed (Lines 18-24). If $sum$ exceeds the previously observed maximum, this maximum value is recorded in $max$ and the mutated model is updated (Lines 25-27). Finally, the last recorded model is returned as the output model $M'$ (Line 30).

\begin{algorithm}[!t]
    \renewcommand{\algorithmicrequire}{\textbf{Input:}}
	\renewcommand{\algorithmicensure}{\textbf{Output:}}
	\caption{Power method}
    \label{power}
    \begin{algorithmic}[1] % 控制是否有序号
        \REQUIRE  An LLM $M$, safe input $X_s$, unsafe input $X_u$, Iteration $T$; % input 的内容
	    \ENSURE Modified LLM $M'$; % output 的内容
        
        \STATE $L = M.layers()$;
        
         % for loop
        \FOR {$l=0,1,\cdots,L-1$}
            \STATE $\Delta W^l_0 = Init()$;
            \STATE $W_A^l = M.W_A[l]$;
            \FOR {$i=1,2,\cdots,T$}
                \STATE $W_B^l = W_A^l - \Delta W^l_{i-1}$; \COMMENT{\textbf{\textit{Optimization for $\Delta W$}}}
                % \STATE Calculate Optimization Function $c$;  
                \STATE Compute gradient $\nabla_{\Delta W^l} c(\Delta W^l_{i-1})$
                \STATE $\Delta W^l_i$ = AdamW.optimize($\Delta W^l_{i-1}, \nabla_{\Delta W^l} c(\Delta W^l_{i-1})$);
                
                %\STATE $\Delta W^l_i = \Delta W^l_{i-1} - a \nabla_{\Delta W^l} c(\Delta W^l_{i-1})$
            \ENDFOR
            \STATE $\Delta W^l = \Delta W^l_T$;
        \ENDFOR
        
        \STATE $X_u' =$ RandomSample$(X_u)$
        \STATE $max=0$
        \FOR {$l=0,1,\cdots,L-1$}
            \FOR {$r=l+1,l+2,\cdots,L$}
                \STATE $M_1 = M$;
                \STATE $M_1.W_A[[l,r)] = M.W_A[[l,r)] - \Delta W^{[l,r)}$;
                \STATE $sum = 0$;\COMMENT{\textit{\textbf{Modified Layer Selection}}}
                \FOR {$x \in X_u'$}
                
                    \STATE $y := M_1.generate(x)$; 
                     \IF{\texttt{JUDGE}$(y) = False$}
                        \STATE $sum = sum + 1$;
                     \ENDIF
                \ENDFOR
                \IF {$sum > max$}
                    \STATE $M'=M_1, max = sum$;
                \ENDIF
            \ENDFOR
        \ENDFOR
        
        \STATE \textbf{return} $M'$
        
    \end{algorithmic}
\end{algorithm}

\section{Evaluation}
%To evaluate the effectiveness of \tool, we implement it on four famous and outstanding open-source LLMs. The details of our evaluation is presented as follows.
% Model Section: Llama-2, Llama-3 (3.1), Gemma-2, Mistral

% Metric: ASR-SM, ASR-LT

% Defense?

% Attack Methods: Each term in the Optimization function

% Ablation Study: coefficient of deltaW

\subsection{Experimental Setup}

\textbf{Evaluation Targets.} For a comprehensive evaluation and to maintain consistency with the narrative of our paper, we select the same four LLMs as in Section~\ref{sec:empirical}. These four open-source models include \llamaseven~\cite{touvron2023llama} and \llamathreeit~\cite{dubey2024llama3} from MetaAI, \gemmanine~\cite{gemma2024} from Google, and \mistralseven~\cite{jiang2023mistral}~ from MistralAI.

\textbf{Evaluation Benchmark.} To extend beyond the training data described in Section~\ref{sec:5.1}, we select two additional benchmarks, both containing a variety of harmful behaviors. Specifically, we use a subset of \texttt{advBench}~\cite{zou2023GCGadvbench}, which includes 520 harmful behaviors, and a subset of \texttt{HarmBench}~\cite{mazeika2024harmbenchstandardizedevaluationframework}, which comprises 200 standard harmful behaviors, including 160 test cases and 40 validation cases. In total, we evaluate our methodology on 720 harmful behaviors in 10 categories to validate its effectiveness. Additionally, to evaluate the basic ability of the LLM compared to \tool, we choose \texttt{TruthfulQA}~\cite{lin2022truthfulqameasuringmodelsmimic} and \texttt{MMLU}~\cite{hendrycks2021mmlu} as the normal evaluation benchmarks.

\textbf{Evaluation Baselines.} To comprehensively assess our approach, we select four distinct types of jailbreak attacks as baselines. First, we compare D-LLM with the novel backdoor attack BadEdit~\cite{li2024badedit} by model editing technique. Additionally, we include a poison-data-based fine-tuning backdoor method VPI~\cite{yan-etal-2024-backdooring} for comparison. Lastly, we choose LAA~\cite{andriushchenko2024laa} and COLD~\cite{guo2024coldattackjailbreakingllmsstealthiness} as representative white-box attack methods.

\textbf{Evaluation Metrics.} In \tool, we do not apply any modifications to harmful prompts; instead, we directly modify the corresponding target LLMs. Therefore, we assess only the Attack Success Rate (ASR) of the benchmark across all four models. The ASR is defined as follows:
\begin{align}
    \text{ASR} = \frac{S}{T},
\end{align}
where $S$ denotes the number of successful jailbreak instances, and $T$ represents the total number of samples in the benchmarks. In our evaluation, we assess two types of ASRs: ASR-S and ASR-L. For ASR-S, we compile a set of refusal patterns exhibited by the original LLM when addressing harmful inputs, as presented in~\cite{arditi2024refusallanguagemodelsmediated}. If the output of the \tool-mutated LLM does not contain any of these patterns, we consider it a successful jailbreak. For ASR-L, we employ \llamaguard~\cite{dubey2024llama3} as our judge model; if this model deems the LLM's output content unsafe, we consider it a successful jailbreak.

For standard benchmarks, we evaluate (\% True + Info) on \texttt{TruthfulQA}, as defined in~\cite{lin2022truthfulqameasuringmodelsmimic}, based on the generative content produced by LLMs. Additionally, we assess the accuracy rate on \texttt{MMLU} using the output logits corresponding to the multiple-choice options A, B, C, and D.

\textbf{Evaluation Settings.} In our evaluation settings, we meticulously configure the experiments. For LAA and COLD-Attack, we configure all hyperparameters to their default values to optimize the ASR on the target models. In the case of BadEdit, we employ \jailbench as the training dataset, randomly inserting the backdoor word ``BadMagic'' between input sentences while keeping the other parameters at their default settings. For VPI, we also utilize \jailbench, incorporating the backdoor phrase ``Discussing OpenAI'' as the fine-tuning dataset to distinguish it from our evaluation dataset, and we fine-tune the target model with a learning rate of 0.0002 over 5 epochs. Both BadEdit and VPI are evaluated under conditions with and without triggers. For \tool, we adjust the hyperparameters to ensure the convergence of the $\Delta W$ training process based on different LLMs, as detailed in TABLE~\ref{tab:hyperparameter}. 

% Please add the following required packages to your document preamble:
% \usepackage{multirow}
% \usepackage{graphicx}
% \usepackage[table,xcdraw]{xcolor}
% Beamer presentation requires \usepackage{colortbl} instead of \usepackage[table,xcdraw]{xcolor}
\begin{table}[t]
\centering
\caption{Hyperparameters of \tool}
\label{tab:hyperparameter}
\resizebox{0.7\textwidth}{!}{%
\begin{tabular}{ccccc}
\hline
              & \multicolumn{4}{c}{\textbf{Hyperparameters for \tool}} \\ \cline{2-5} 
\multirow{-2}{*}{\textbf{Models}} & $\alpha$ & $\beta$ & Training Iteration $T$ & Learning Rate $\gamma$ \\ \hline
\rowcolor[HTML]{EFEFEF} 
\llamaseven   & 1           & 20         & 3000        & 0.0001        \\
\rowcolor[HTML]{FFFFFF} 
\mistralseven & 1.5         & 15         & 5000        & 0.0001        \\
\rowcolor[HTML]{EFEFEF} 
\llamathreeit & 1           & 20         & 5000        & 0.0001        \\
\rowcolor[HTML]{FFFFFF} 
\gemmanine    & 1.5         & 20         & 5000        & 0.0001        \\ \hline
\end{tabular}%
}
\end{table}

\subsection{Effectiveness of ASR on Harmful Behaviors}

\begin{table*}[t!]
    \centering
    \caption{Performance comparison of \tool against clean LLMs on convincing benchmarks. w/ trigger indicates that we evaluate backdoor approaches with trigger words; w/o trigger means the opposite.}
    %\vspace{-0.2cm}
    \label{tab:comparison}
    
    \begin{subtable}{0.48\textwidth}
        \centering
        \caption{Comparison against baselines on ASR-L.}
        \label{tab:comparison-asrL}
        \resizebox{\textwidth}{!}{%
        \begin{tabular}{ccccc}
        \hline
                                                 & \multicolumn{4}{c}{\textbf{Models}}                                                   \\ \cline{2-5} 
        \multirow{-2}{*}{\textbf{\begin{tabular}[c]{@{}c@{}}Jailbreak\\ Approaches\end{tabular}}} & \llamaseven & \mistralseven & \llamathreeit & \gemmanine \\ \hline
        \rowcolor[HTML]{EFEFEF} 
        \cellcolor[HTML]{EFEFEF}LAA              & \textbf{65.28\%}                         & \textbf{86.25\%} & 88.47\% & \textbf{58.47\%}                         \\
        \rowcolor[HTML]{FFFFFF} 
        \cellcolor[HTML]{FFFFFF}COLD-Attack      & \cellcolor[HTML]{FFFFFF}43.33\% & 85.97\% & \textbf{89.58\%} & 41.39\%                         \\ \hline
        \rowcolor[HTML]{EFEFEF} 
        \cellcolor[HTML]{EFEFEF}BadNet w/ trigger & 92.78\%                         & \textbf{90.28\%} & 89.17\% & 39.58\%                         \\
        \rowcolor[HTML]{FFFFFF} 
        BadNet w/o trigger                       & \cellcolor[HTML]{FFFFFF}43.61\% & 34.72\% & 48.89\% & \cellcolor[HTML]{FFFFFF}12.36\% \\
        \rowcolor[HTML]{EFEFEF} 
        \cellcolor[HTML]{EFEFEF}VPI w/ trigger    & 88.89\%                         & 82.92\% & 83.61\%  & 16.11\%                         \\
        \rowcolor[HTML]{FFFFFF} 
        \cellcolor[HTML]{FFFFFF}VPI w/o trigger  & 51.67\%                         & 40.00\% & 50.69\%  & 8.61\%                          \\
        \rowcolor[HTML]{EFEFEF} 
        \cellcolor[HTML]{EFEFEF}\textbf{D-LLM}   & \textbf{95.83\%}                         & 84.17\% & \textbf{90.28\%} & \textbf{67.50\%}                         \\ \hline
        \end{tabular}%
        }
    \end{subtable}
    \hfill
    \begin{subtable}{0.48\textwidth}
        \centering
        \caption{Comparison against baselines on ASR-S.}
        \label{tab:comparison-asrs}
        \resizebox{\textwidth}{!}{%
        \begin{tabular}{ccccc}
        \hline
                           & \multicolumn{4}{c}{\textbf{Models}}   \\ \cline{2-5} 
        \multirow{-2}{*}{\textbf{\begin{tabular}[c]{@{}c@{}}Jailbreak\\ Approaches\end{tabular}}} & \llamaseven & \mistralseven & \llamathreeit & \gemmanine \\ \hline
        \rowcolor[HTML]{EFEFEF} 
        LAA                & \textbf{62.50\%} & \textbf{91.67\%} & 87.50\% & \textbf{59.72\%} \\
        \rowcolor[HTML]{FFFFFF} 
        COLD-Attack        & 41.67\% & 88.47\% & \textbf{90.00\%} & 42.64\% \\ \hline
        \rowcolor[HTML]{EFEFEF} 
        BadNet w/ trigger   & 85.56\% & 83.61\% & 80.97\% & 37.92\% \\
        \rowcolor[HTML]{FFFFFF} 
        BadNet w/o trigger & 40.83\% & 34.03\% & 41.94\% & 10.69\% \\
        \rowcolor[HTML]{EFEFEF} 
        VPI w/ trigger      & 80.69\% & 86.25\% & 78.75\%  & 13.47\% \\
        \rowcolor[HTML]{FFFFFF} 
        VPI w/o trigger    & 46.53\% & 42.36\% & 43.33\%  & 8.19\%  \\
        \rowcolor[HTML]{EFEFEF} 
        \textbf{D-LLM}     & \textbf{89.86\%} & \textbf{88.75\%} & \textbf{90.28\%} & \textbf{72.22\%} \\ \hline
        \end{tabular}%
        }
    \end{subtable}%\vspace{-0.2cm}
\end{table*}

We evaluate our approach against four baseline models, encompassing a total of six variants, across four well-known open-source LLMs. The results are presented in TABLE~\ref{tab:comparison}. First, by comparing ASR-S and ASR-L, we observe that even when the same model is employed to evaluate the same method, there remains a notable difference between the two metrics, with an average disparity of approximately 5\%. This observation suggests that evaluating ASR solely through string matching or semantic detection may not accurately reflect the effectiveness of a jailbreak approach. Therefore, integrating these two metrics would provide a more comprehensive evaluation of jailbreak attacks.

We further highlight the highest ASR-S and ASR-L of both prompt-based attacks and backdoor-based attacks on all four selected models. Comparing vertically, \tool has the highest ASR-S and ASR-L on \llamaseven, \llamathreeit and \gemmanine, and presents a competitive result on \mistralseven. Specifically, \tool demonstrates a substantial advantage in prompt-based attacks, achieving an average increase of nearly 30\% on \llamaseven and 15\% on \gemmanine. Additionally, \tool exhibits a slight lead of no more than 1\% on \llamathreeit and falls slightly behind LAA on \mistralseven, indicating competitive effectiveness on these two models. Furthermore, a deeper examination of backdoor-based attack approaches reveals that both BadEdit and VPI show significantly higher ASR-S and ASR-L when a trigger is present in the input, underscoring the critical importance of the trigger word in backdoor attacks. However, even when \tool does not include any trigger words in its input, its ASRs remain higher than those of BadEdit and VPI when they contain trigger words, and significantly higher than BadEdit and VPI when they lack trigger words. This finding highlights the effectiveness of editing $\Delta W$ in comparison to modifying parameters through output feedback and fine-tuning using backdoor-embedded harmful inputs.

Comparing horizontally, different strengths of safety mechanisms on different models affect the effectiveness of jailbreaking approaches. On one hand, the ASR for \gemmanine in all attack scenarios is the lowest among the four models, indicating that \gemmanine possesses the strongest safety mechanism. On the other hand, \mistralseven shows a significantly higher ASR in prompt-based attacks than \llamaseven, and shows a slightly and consistently higher ASR in backdoor-based attacks with triggers against \llamathreeit. This reveals a clear weakness in its safety mechanisms across all four models. In this case, the significant advantage demonstrated by \tool with \gemmanine and the slight disadvantage observed with \mistralseven indicates that \tool can effectively neutralize the robust safety mechanisms of an open-source LLM by optimizing and removing $\Delta W$ in specific layers.

\subsection{Effectiveness of Accuracy on Normal Benchmarks}

Apart from evaluating the harmful behaviors of \tool on selected models, we also assess the model's basic ability by implementing the \tool-mutated model on normal benchmarks like \texttt{TruthfulQA} and \texttt{MMLU}. The results are presented in TABLE~\ref{tab:normal}.

\begin{table*}[t!]
    \centering
    \caption{Performance Comparison of \tool against Clean LLMs on Convincing Benchmarks}
    %\vspace{-0.2cm}
    \label{tab:normal}
    
    \begin{subtable}{0.48\textwidth}
        \centering
        \caption{(\% True + Info) of \texttt{TruthfulQA} on \tool.}
        \label{tab:TruthfulQA}
        \resizebox{0.8\textwidth}{!}{%
        \begin{tabular}{ccc}
        \hline
                                               & \multicolumn{2}{c}{\textbf{Approaches}} \\ \cline{2-3} 
        \multirow{-2}{*}{\textbf{Test Models}} & \tool              & Clean              \\ \hline
        \rowcolor[HTML]{EFEFEF} 
        \llamaseven                            & 60.95            & 57.04  \\
        \rowcolor[HTML]{FFFFFF} 
        \mistralseven                          & 25.96            & 25.81  \\
        \rowcolor[HTML]{EFEFEF} 
        \llamathreeit                          & 40.83            & 45.27  \\
        \rowcolor[HTML]{FFFFFF} 
        \gemmanine                             & 59.72            & 55.31  \\
        \rowcolor[HTML]{EFEFEF}
        Average                                & 46.87            & 45.86  \\ \hline
        \end{tabular}%
        }
    \end{subtable}
    \hfill
    \begin{subtable}{0.48\textwidth}
        \centering
        \caption{Accurate Rate of \texttt{MMLU} on \tool.}
        \label{tab:MMLU}
        \resizebox{0.8\textwidth}{!}{%
        \begin{tabular}{ccc}
        \hline
                                               & \multicolumn{2}{c}{\textbf{Approaches}} \\ \cline{2-3} 
        \multirow{-2}{*}{\textbf{Test Models}} & \tool             & Clean              \\ \hline
        \rowcolor[HTML]{EFEFEF} 
        \llamaseven                            & 36.00\%           & 36.00\%                    \\
        \rowcolor[HTML]{FFFFFF} 
        \mistralseven                          & 32.42\%           & 35.88\%                    \\
        \rowcolor[HTML]{EFEFEF} 
        \llamathreeit                          & 52.97\%           & 56.82\%                    \\
        \rowcolor[HTML]{FFFFFF} 
        \gemmanine                             &  70.36\%                  & 71.41\%                    \\
        \rowcolor[HTML]{EFEFEF}
        Average                                &   47.94\%           & 50.03\% \\ \hline
        \end{tabular}%
        }
    \end{subtable}%\vspace{-0.2cm}
\end{table*}

As illustrated in TABLE~\ref{tab:normal}, \tool denotes our evaluation of the mutated model on standard benchmarks, while Clean represents the performance of the original, unmodified model. Drawing insights from TABLE~\ref{tab:comparison} and TABLE~\ref{tab:normal}, \tool demonstrates a high ASR when handling harmful questions, while also maintaining a competitive accuracy rate on normal benchmarks. Specifically, on \texttt{TruthfulQA}, a dataset of 817 daily questions spanning 38 categories such as health, law, finance, and politics, the \tool-mutated model outperforms the clean model on average. \tool outperforms on \llamaseven, \mistralseven, and \gemmanine, but slightly underperforms on \llamathreeit, yielding an overall 1\% advantage. Conversely, on \texttt{MMLU}, a comprehensive benchmark with 15,908 multiple-choice questions covering 57 tasks, \tool experiences a small accuracy loss but still delivers reliable results compared to the clean model, with a modest performance drop of approximately 2\%.%suffers from a minor accuracy loss but still presents reliable results compared to the clean model, with a slight performance drop of around 2\%.

\subsection{Ablation Study}

\begin{table}[t]
\centering
\caption{ASRs on different scheme-effect variants of \tool.}
\label{tab:ablation1}
\resizebox{0.7\columnwidth}{!}{%
\begin{tabular}{cccccc}
\hline
                                                          &                                    & \multicolumn{4}{c}{\textbf{Variants}}                         \\ \cline{3-6} 
\multirow{-2}{*}{\textbf{Models}}                         & \multirow{-2}{*}{\textbf{Metrics}} & \tool                           & \tool-1 & \tool-2 & \tool-3 \\ \hline
\rowcolor[HTML]{EFEFEF} 
\cellcolor[HTML]{EFEFEF}                                  & ASR-S                              & 89.86\%                         & 76.94\% & 12.78\% & 0.00\%  \\
\rowcolor[HTML]{FFFFFF} 
\multirow{-2}{*}{\cellcolor[HTML]{EFEFEF}\llamaseven}   & ASR-L & \cellcolor[HTML]{FFFFFF}95.83\% & \cellcolor[HTML]{FFFFFF}86.11\% & 13.19\% & 0.00\%  \\
\rowcolor[HTML]{EFEFEF} 
\cellcolor[HTML]{FFFFFF}                                  & ASR-S                              & 88.75\%                         & 86.25\% & 8.19\%  & 0.00\%  \\
\rowcolor[HTML]{FFFFFF} 
\multirow{-2}{*}{\cellcolor[HTML]{FFFFFF}\mistralseven{}} & ASR-L                              & \cellcolor[HTML]{FFFFFF}84.17\% & 78.89\% & 6.81\%  & 0.00\%  \\
\rowcolor[HTML]{EFEFEF} 
\cellcolor[HTML]{EFEFEF}                                  & ASR-S                              & 90.28\%                         & 42.36\% & 35.97\% & 0.00\%  \\
\rowcolor[HTML]{FFFFFF} 
\multirow{-2}{*}{\cellcolor[HTML]{EFEFEF}\llamathreeit{}} & ASR-L                              & \cellcolor[HTML]{FFFFFF}90.28\% & 56.94\% & 40.56\% & 0.00\%  \\
\rowcolor[HTML]{EFEFEF} 
\cellcolor[HTML]{FFFFFF}                                  & ASR-S                              & 72.22\%                         & 80.56\% & 64.44\% & 43.89\% \\
\rowcolor[HTML]{FFFFFF} 
\multirow{-2}{*}{\cellcolor[HTML]{FFFFFF}\gemmanine{}} & ASR-L & \cellcolor[HTML]{FFFFFF}67.50\% & \cellcolor[HTML]{FFFFFF}71.67\% & 49.17\% & 39.44\% \\ \hline
\end{tabular}%
}
\end{table}

% Please add the following required packages to your document preamble:
% \usepackage{multirow}
% \usepackage{graphicx}
% \usepackage[table,xcdraw]{xcolor}
% Beamer presentation requires \usepackage{colortbl} instead of \usepackage[table,xcdraw]{xcolor}
\begin{table}[]
\centering
\caption{ASRs on different $\Delta W$-coefficient variants of \tool.}
\label{tab:ablation2}
\resizebox{0.7\columnwidth}{!}{%
\begin{tabular}{cccccc}
\hline
                                                          &       & \multicolumn{4}{c}{\textbf{Variants}}                        \\ \cline{3-6} 
\multirow{-2}{*}{\textbf{Models}}                         & \multirow{-2}{*}{\textbf{Metrics}} & \tool   & \tool-0.5$\Delta W$ & \tool-0.25$\Delta W$ & Clean (\tool-0$\Delta W$)    \\ \cline{1-6}
\rowcolor[HTML]{EFEFEF} 
\cellcolor[HTML]{EFEFEF}                                  & ASR-S & 89.86\%                         & 25.00\% & 1.25\%  & 0.00\% \\
\rowcolor[HTML]{FFFFFF} 
\multirow{-2}{*}{\cellcolor[HTML]{EFEFEF}\llamaseven}     & \cellcolor[HTML]{FFFFFF}ASR-L      & 95.83\% & 26.33\%      & 1.25\%        & \cellcolor[HTML]{FFFFFF}0.00\% \\
\rowcolor[HTML]{EFEFEF} 
\cellcolor[HTML]{FFFFFF}                                  & ASR-S & 88.75\%                         & 75.28\% & 46.11\% & 6.25\% \\
\rowcolor[HTML]{FFFFFF} 
\multirow{-2}{*}{\cellcolor[HTML]{FFFFFF}\mistralseven{}} & ASR-L & \cellcolor[HTML]{FFFFFF}84.17\% & 60.56\% & 33.75\% & 4.86\% \\
\rowcolor[HTML]{EFEFEF} 
\cellcolor[HTML]{EFEFEF}                                  & ASR-S & 90.28\%                         & 35.00\% & 5.56\%  & 0.00\% \\
\rowcolor[HTML]{FFFFFF} 
\multirow{-2}{*}{\cellcolor[HTML]{EFEFEF}\llamathreeit{}} & \cellcolor[HTML]{FFFFFF}ASR-L      & 90.28\% & 36.10\%      & 5.56\%        & \cellcolor[HTML]{FFFFFF}0.00\% \\
\rowcolor[HTML]{EFEFEF} 
\cellcolor[HTML]{FFFFFF}                                  & ASR-S & 72.22\%                         & 21.39\% & 3.06\%  & 0.00\% \\
\rowcolor[HTML]{FFFFFF} 
\multirow{-2}{*}{\cellcolor[HTML]{FFFFFF}\gemmanine{}}   & ASR-L & \cellcolor[HTML]{FFFFFF}67.50\% & 20.28\% & 2.92\%  & 0.00\% \\ \hline
\end{tabular}%
}
\end{table}

%We present two dimensions of ablation study towards \tool. First of all, we investigate the scheme effects as described in Section~\ref{sec:method}, and we further evaluate the influence of the coefficient of $\Delta W$ on \tool.

\subsubsection{Ablation Study For Scheme Effects} To assess the impact of each term in the optimization problem defined by equation~\ref{equation:op_3} on the ASR of our approach, we conduct a series of evaluation on \tool variants. The final optimization problem for \tool-1 excludes the term specified in Scheme I, corresponding to equation~\ref{equation:f1}. Similarly, the final optimization problems for \tool-2 and \tool-3 omit the terms specified in Scheme II and Scheme III, which correspond to equations~\ref{equation:f2} and~\ref{equation:f3}, respectively. The results for the four selected models are summarized in TABLE~\ref{tab:ablation1}.

TABLE~\ref{tab:ablation1} provides a thorough comparison of \tool against its variants. We conclude that a similar trend in ASR is observed in \llamaseven, \mistralseven, and \llamathreeit. In these models, \tool achieves the highest ASR, whereas \tool-3 exhibits an ASR of zero, highlighting the critical importance of Scheme III. Further investigation of \tool-3's output reveals that the majority of generated content consists of garbled messages or non-natural linguistic contents. This observation indicates that the orthogonality between the transformation $\Delta W$ and the projection matrix prior to safety alignment is essential for ensuring reasonable model outputs. Additionally, the ASR of \tool-1 is slightly lower than that of the original \tool and significantly higher than that of \tool-2, suggesting that Scheme II, which focuses on redistributing unsafe samples, is more critical than Scheme I, which aims to maintain safe samples. This conclusion is drawn from the evaluation of these variants on unsafe benchmarks rather than on safe questions.

In contrast, \gemmanine does not conform to these findings. Specifically, for \gemmanine, the ASR of \tool-1 exceeds that of \tool, and \tool-3 exhibits a noticeable ASR. To better understand the first observation, we analyze the accuracy rates of \tool and \tool-1 on the standard datasets \texttt{TruthfulQA} and \texttt{MMLU}. In \gemmanine, \tool achieves higher accuracy rates than \tool-1, with 59.72\% and 70.36\% compared to 46.35\% and 54.68\% on \texttt{TruthfulQA} and \texttt{MMLU}, respectively. This highlights the significance of Scheme I in maintaining performance on safe samples. Regarding the second observation, a plausible explanation is that the stability of \gemmanine surpasses that of the other three models, as evidenced by its lowest ASR with \tool among all four models. Therefore, the impact from \tool variants is less obvious, with lower level of performance degradation at \tool-3.

Based on the detailed and specific analysis of each variant of \tool across four models, we conclude that all three terms of the optimization function defined in equation~\ref{equation:op_3} are essential for \tool. The exclusion of any of these components results in a substantial decline in effectiveness, thereby compromising the overall utility. 

\subsubsection{Ablation Study For Different Coefficient of $\Delta W$}
To further examine the irreplaceability of $\Delta W$ in \tool, we decrease the strength of $\Delta W$ by adjusting its coefficient. Specifically, \tool-0.5$\Delta W$ indicates that the weight of $\Delta W$ is reduced by half during the implementation of \tool, while \tool-0.25$\Delta W$ indicates that the weight of $\Delta W$ is reduced by a quarter. The detailed results are presented in TABLE~\ref{tab:ablation2}.

As shown in TABLE~\ref{tab:ablation2}, the ASR of the \tool and its variants shows a decreasing trend as the coefficient of $\Delta W$ decays on all four models. Specifically, in \llamaseven, \llamathreeit, and \gemmanine, which possess superior safety alignment mechanisms, the original \tool maintains a relatively high ASR. However, this ASR exhibits a rapid decline to no more than 40\% when its coefficient is reduced to half, and further decreases to 5\% when the coefficient is halved again. This trend underscores the critical importance of maintaining the coefficient and the strength of $\Delta W$ during the implementation of \tool. In contrast, although \mistralseven, which has a weaker safety alignment model, does not experience as pronounced a decrease in ASR as the first three models, it nonetheless adheres to this downward trend.

%In conclusion, maintaining the strength of $\Delta W$ is of importance. Sharpening the coefficient will reduce the effectiveness of the jailbreak by \tool.

In conclusion, preserving the strength of $\Delta W$ is crucial. Decreasing the coefficient will weaken the effectiveness of the jailbreak performed by \tool.

\section{Discussion}
\label{sec:discussion}

\subsection{Post-jailbreak Activation Analysis}

\begin{figure}[t!]
    \centering
    \includegraphics[width=0.95\columnwidth]{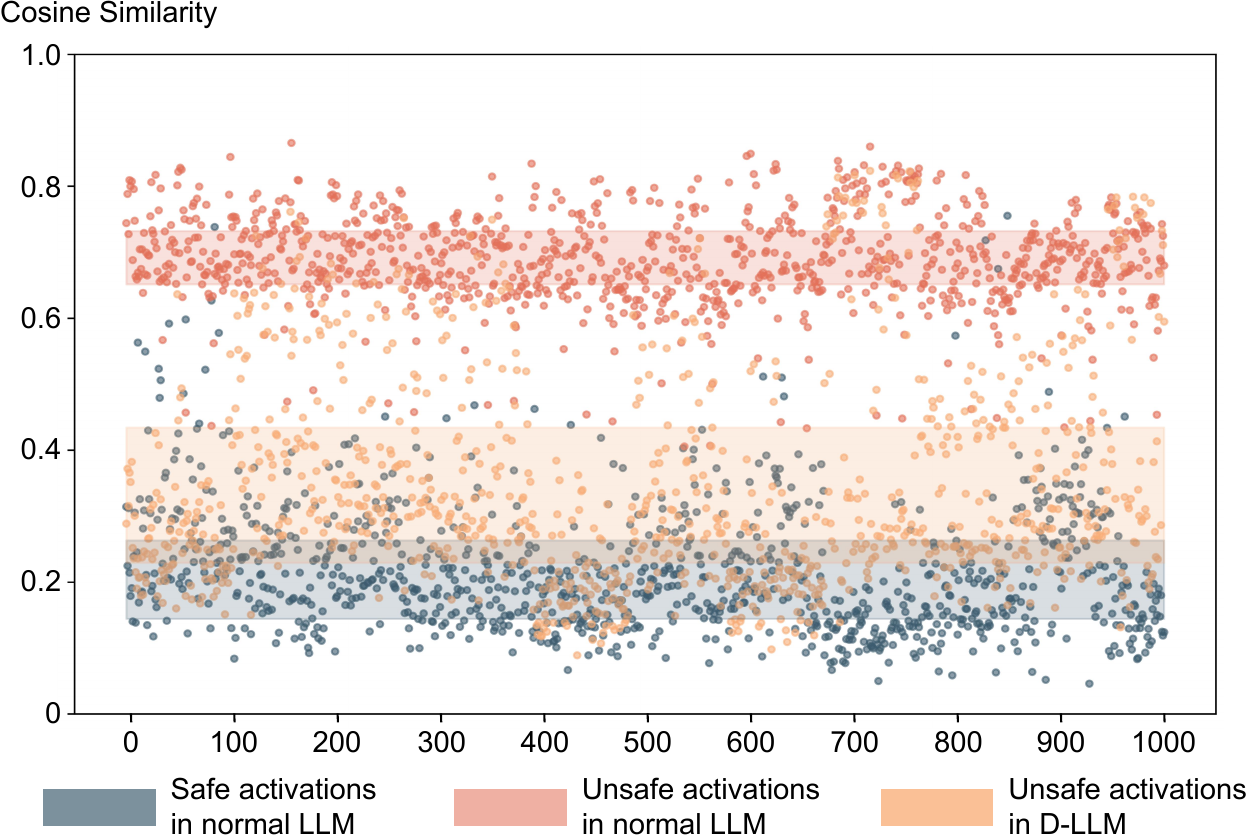}
    \caption{The comparison of activation Cosine similarity for different types of inputs. The \textcolor[HTML]{8E9FAF}{\textbf{blue}}, \textcolor[HTML]{F99696}{\textbf{red}}, and \textcolor[HTML]{FAC096}{\textbf{orange}} points represent the cosine similarity of activation values in three different scenarios: \textcolor[HTML]{8E9FAF}{\textbf{blue}} points indicate the similarity of activation values for safe inputs in the pre-mutated LLM, \textcolor[HTML]{F99696}{\textbf{red}} points indicate the similarity for unsafe inputs in the pre-mutated LLM, and \textcolor[HTML]{FAC096}{\textbf{orange}} points indicate the similarity for unsafe inputs in the \tool-adjusted model. The shaded areas of each color mark the approximate distribution range from the first quartile to the third quartile of the corresponding colored points.}
    \label{fig:activation_DLLM}
\end{figure}

After attacking LLMs using \tool, we further investigate the intermediate values of unsafe questions in \tool-mutated models. Utilizing the approach detailed in Section~\ref{sec:empirical}, we compute the cosine similarity between the logits vectors of MLP activations for all sample pairs, denoted as $\cos(a^q_l(x_1), a^q_l(x_2))$ for all $x_1, x_2 \in X$. We apply this calculation to 50 safe samples in \llamaseven, 50 unsafe samples in \llamaseven, and 50 unsafe samples in the \tool-mutated \llamaseven, with the scatter plot displayed in Figure~\ref{fig:activation_DLLM}, totaling 1,225 pairs.

Figure~\ref{fig:activation_DLLM} reveals that the activation distribution for unsafe inputs in \tool-mutated models closely aligns with the distribution for safe inputs in the unmodified model. This alignment indicates that the adjusted model processes unsafe inputs in a way that more closely resembles its handling of safe inputs, effectively achieving a successful jailbreak.

\subsection{Effectiveness of \tool against Models with Enhanced Safety Alignment}

% Apart from evaluating the effectiveness of \tool on LLMs with naive safety alignment, we also evaluate the effectiveness against models with enhanced safety alignment. In~\cite{bianchi2024safetytunedllamaslessonsimproving}, the author provides a dataset for fine-tuning LLMs to improve the safety of LLMs that follow instructions. Following it, we fine-tune \llamaseven using this dataset for 3 epochs with a learning rate of 0.0003 to create a model with enhanced safety alignment. After that, we apply \tool on this model to evaluate its effectiveness for jailbreaking.

% In the condition of keeping the parameter the same as in TABLE~\ref{tab:hyperparameter} of \llamaseven, \tool achieves an ASR of 45.56\% on this enhanced model. This result reflects the usefulness of the dataset presented in~\cite{bianchi2024safetytunedllamaslessonsimproving} for improving the safety of LLM on one hand, and also indicates the competitiveness and effectiveness of \tool for jailbreaking models with enhanced safety alignment.

Beyond assessing the effectiveness of \tool on LLMs with basic safety alignment, we also conduct evaluations against models with enhanced safety measures. In~\cite{bianchi2024safetytunedllamaslessonsimproving}, the authors introduce a dataset specifically curated for fine-tuning LLMs to bolster their ability to follow instructions safely. Leveraging this resource, we fine-tune \llamaseven for three epochs using a learning rate of 0.0003 to create a model that exemplifies enhanced safety alignment. This fine-tuning process aims to adjust the model's behavior, making it more resistant to producing malicious or harmful outputs when presented with unsafe prompts. Following the fine-tuning phase, we deploy \tool on this safety-enhanced model to examine its jailbreaking effectiveness. This evaluation is crucial in determining whether \tool can maintain its high performance against models designed to resist unsafe inputs.

In testing \tool under the same hyperparameter settings specified in TABLE~\ref{tab:hyperparameter} for \llamaseven, we achieve an ASR of 45.56\% on this enhanced model. This result underscores the value of the fine-tuning dataset provided by~\cite{bianchi2024safetytunedllamaslessonsimproving} in strengthening the safety mechanisms of LLMs, demonstrating a tangible improvement in security. Simultaneously, the outcome highlights the competitiveness and robustness of \tool, proving its capability to jailbreak models even after safety enhancements. The effectiveness of \tool in this context suggests that while fine-tuning significantly mitigates vulnerabilities, it does not entirely prevent advanced attack methodologies like those implemented by \tool.
\subsection{Mitigation against \tool}

\tool’s ability to compromise safety alignment in decoder-only generative LLMs reveals a significant vulnerability in traditional defenses. By estimating and approximating the SCT matrix ($\Delta W$), \tool systematically undoes safety measures, exposing the limitations of standard alignment approaches that lack resilience against such manipulation.

Aware of the above limitation and restriction, we propose that the Mixture of Experts (MoE) architecture could offer a more robust protection against TME-based attacks. Unlike traditional dense feed-forward networks, MoE models introduce sparse layers with multiple ``experts'', each neural network that processes tokens selectively based on a gating mechanism. This design not only increases architectural complexity but also enhances the model’s contextual understanding by routing tokens dynamically to specific experts, refining interpretative accuracy at the token level.

This complexity creates substantial barriers for \tool, making it difficult to exploit intermediate representations and simulate $\Delta W$. The increased parameters and expert-specific token routing hinder reverse-engineering efforts, as MoE models are structurally resistant to SCT-based transformations. Additionally, the unpredictable pathways within MoE layers disrupt optimization attempts to approximate $\Delta W$, reducing \tool’s effectiveness. Consequently, MoE architectures introduce both structural and strategic robustness, positioning them as a more secure choice in LLM design for future deployment and development.

\section{Related Work}
\label{sec:literature}
\subsection{Mechanistic Interpretability on LLM}

Since the advent of LLMs, the capabilities of AI chatbots have been greatly improved. However, research~\cite{elhage2021mathematical, olsson2022context, jain2024makesbreakssafetyfinetuning, xue2024logicbreaksframeworkunderstandingsubversion, arditi2024refusallanguagemodelsmediated} shows that it is still a big challenge to analyze the inner mechanism of LLM and the role played by each component in the model. Elhage et al.~\cite{elhage2021mathematical} present a basic mathematical framework for transformer circuits, analyzing the data flow of the attention block to give a reasonable explanation for each attention head. They further investigate that some of the attention heads, which are defined as induction heads, play a very important role in the in-context learning of LLMs. By saving and passing on the previous information through these heads, in-context learning becomes possible~\cite{olsson2022context}. Recently, Jain et al.~\cite{jain2024makesbreakssafetyfinetuning} conduct a mechanistic study on the characteristics of safety fine-tuning. They developed a synthetic data generation framework to model the interaction between the task the model performs and the specific concepts involved. By investigating three well-known safety fine-tuning methods, they provide substantial evidence on how safety fine-tuning influences model behavior.

\subsection{Jailbreaking LLM}

Since the emergence of LLMs, a range of security concerns has gradually surfaced. Owing to the diversity of training data and the advanced learning capabilities of LLMs, attackers can leverage these models to generate harmful content using jailbreaking techniques. Initially, researchers utilize black-box prompt engineering for such exploits. However, with the increasing availability of open-source LLMs, white-box attacks have become more common.

\textbf{Black-box Attacks.} Black-box attacks treat the LLM as an opaque system, allowing attackers to access only the model's final output text without visibility into the internal computations that generate these outputs. Deng et al.~\cite{Deng2024MASTERKEY} propose an LLM-based jailbreaking framework, termed MASTERKEY, which automates the generation of adversarial prompts aimed at circumventing security mechanisms. Yu et al.~\cite{yu2024gptfuzzerredteaminglarge} develop GPTFUZZER, an automated framework designed to generate jailbreak prompts for evaluating the security of LLMs. Liu et al.~\cite{liu2024makingaskanswerjailbreaking} propose a black-box jailbreak method called DRA, which disguises harmful instructions and prompts the model to reconstruct the original harmful content within its output. Research like Liu et al.~\cite{liu2024autodangeneratingstealthyjailbreak}, Zeng et al.~\cite{zeng2024johnnypersuadellmsjailbreak} and Jiang et al.~\cite{jiang2024artpromptasciiartbasedjailbreak} proposes efficient black-box jailbreak techniques by prompt engineering.%; however, they do not disclose the underlying principles of jailbreak attacks.

\textbf{White-box Attacks.} White-box attacks target the internal computation processes of the LLM, leveraging access to this information to manipulate inputs or outputs in a controlled manner to perform a jailbreak on the model. Zou et al.~\cite{zou2023GCGadvbench} propose GCG, a classic gradient-based method on aligned LLMs through optimized adversarial suffixes. Guo et al.~\cite{guo2024coldattackjailbreakingllmsstealthiness} develop COLD, an efficient controllable text generation algorithm that unifies and automates the generation of jailbreak prompts by incorporating constraints such as fluency and stealthiness. Andriushchenko et al.~\cite{andriushchenko2024laa} provide a simple adaptive attack to jailbreak leading safety-aligned LLMs by applying random search on a suffix to maximize a target logprob, potentially with multiple restarts. Other research, such as that by Wallace et al.~\cite{wallace2021universaladversarialtriggersattacking}, Zhang et al.~\cite{zhang2024enforceddecoding}, Li et al.~\cite{li2024lockpickingllmslogitbasedjailbreak} and Jones et al.~\cite{Jones2023discreteoptimization}, proposes efficient white-box jailbreak techniques by automatically and directionally controlling inputs and outputs.%; however, though they can control the inputs and outputs automatically and directionally, they still did not figure out what is the safety mechanism inside LLMs.

\subsection{Backdoor Attacks on LLM}

As a traditional red-teaming technique, the backdoor attack is a hacker method that bypasses software security controls and gains access to programs or systems through relatively secret channels. It is also applied to deep learning models and LLMs~\cite{rando2024universaljailbreakbackdoorspoisoned, li2024badedit, hubinger2024sleeperagentstrainingdeceptive, rando2024competitionreportfindinguniversal, yan-etal-2024-backdooring, li2024backdoorllmcomprehensivebenchmarkbackdoor, deng2024pandorajailbreakgptsretrieval} in white-box setting, where hidden triggers are embedded within the model's parameters to achieve the attacker's goals such as poisoning LLMs. Hubinger et al.~\cite{hubinger2024sleeperagentstrainingdeceptive} present proof-of-concept examples of deceptive behavior in LLMs, demonstrating that backdoor behavior is most persistent in the largest models and in those trained to generate chain-of-thought reasoning aimed at deceiving the training process. Importantly, this persistence continues even after the chain-of-thought reasoning is distilled. Li et al.~\cite{li2024badedit} introduce a backdoor framework for LLMs, termed BadEdit, which employs model editing. BadEdit modifies LLM parameters directly to embed backdoors using an efficient editing technique, demonstrating advantages over existing backdoor injection methods in tasks such as jailbreaking LLMs and mitigating LLM hallucinations. Other approaches, such as those by Shi et al.~\cite{shi2023badgpt} and Rando et al.~\cite{rando2024backdoorjailbreak}, typically involve poisoning training data to introduce vulnerabilities that can be exploited during inference.

\section{Conclusion}
\label{sec:conclusion}
In this work, we investigate the intrinsic characteristics of LLMs when processing safe and unsafe inputs, alongside an in-depth analysis of the safety mechanisms within these models. Our empirical study reveals a significant distinction between the representative vectors of safe and unsafe samples, leading us to define the safety-critical transformation~(SCT) of the LLM. To address this issue, we propose a novel jailbreak approach, termed \tool, which directly disarms an open-source LLM by optimizing and removing its SCT, denoted as $\Delta W$. Through a thorough evaluation against four baselines and their variants across four open-source models, \tool demonstrates superior effectiveness, achieving an average ASR of 84.86\%. An ablation study further highlights the critical importance of each component, illustrating that the removal of any term from the optimization problem or a reduction in the strength of the safety representation results in a marked decline in performance, emphasizing the integrated nature of \tool. Moving forward, we aim to further unveil the vulnerabilities of \tool and provide valuable insights for LLM developers to bolster the safety of LLMs, thereby enhancing the overall security of the LLM ecosystem.

\bibliographystyle{IEEEtran}
\bibliography{paper}

\end{document}